\documentclass[twoside,11pt]{article}

\usepackage{palatino}
\usepackage{setspace}
\usepackage{amsthm}
 \usepackage[margin=1.00in, centering]{geometry}
\usepackage{amsfonts}
\usepackage{amsmath}
\usepackage{bigints}
\usepackage{mathtools}
\usepackage{float}
\usepackage[normalem]{ulem}
\usepackage{graphicx}
\usepackage[hyphens]{url}
\usepackage[table, dvipsnames]{xcolor}
\definecolor{darkblue}{rgb}{0.0,0.0,0.6}
\RequirePackage[colorlinks,citecolor=darkblue,urlcolor=darkblue, linkcolor=darkblue]{hyperref} 
\hypersetup{breaklinks=true}
\usepackage{subcaption}
\usepackage{algorithm}
\usepackage{algpseudocode}

\usepackage{multirow}
\usepackage{tikz}
\usepackage{cancel}
\newcommand{\sbt}{\,\begin{picture}(-1,1)(-1,-1)\circle*{3}\end{picture}\ }
\newcommand{\cbglinv}{C_{\mathrm{BGL}}^{-1}}
\newcommand{\cghsinv}{C_{\mathrm{GHS}}^{-1}}
\usepackage{natbib}
\defcitealias{chib1995marginal}{Chib's (1995)}
\defcitealias{wang2012bayesian}{Wang's (2012)}
\allowdisplaybreaks
\newcommand{\Ind}{1\!\mathrm{l}}

\newcommand{\kron}{\otimes}

\renewcommand{\a}{{\bf a}}

\newcommand{\f}{{\bf f}}
\newcommand{\g}{{\bf g}}

\newcommand{\s}{{\bf s}}

\renewcommand{\v}{{\bf v}}

\newcommand{\sv}{{\mathrm{v}}}
\newcommand{\sw}{{\mathrm{w}}}
\newcommand{\sg}{{\mathrm{g}}}

\newcommand{\y}{{\bf y}}

\newcommand{\A}{{\bf A}}

\newcommand{\C}{{\bf C}}

\newcommand{\G}{{\bf G}}

\newcommand{\I}{{\bf I}}

\renewcommand{\S}{{\bf S}}

\newcommand{\U}{{\bf U}}
\newcommand{\V}{{\bf V}}
\newcommand{\W}{{\bf W}}

\newcommand{\bbeta}{\boldsymbol{\beta}}
\newcommand{\btau}{\boldsymbol{\tau}}

\newcommand{\bzeta}{\boldsymbol{\zeta}}

\newcommand{\btheta}{\boldsymbol{\theta}}

\newcommand{\bGamma}{\mathbf{\Gamma}}

\newcommand{\bmu}{\boldsymbol{\mu}}

\newcommand{\bomega}{\boldsymbol{\omega}}
\newcommand{\bOmega}{\boldsymbol{\Omega}}

\newcommand{\wgamma}{\widetilde{\gamma}}
\newcommand{\womega}{\widetilde{\omega}}

\newcommand{\wbomega}{\boldsymbol{\widetilde{\omega}}}
\newcommand{\wbtheta}{\boldsymbol{\widetilde{\theta}}}
\newcommand{\wbbeta}{\boldsymbol{\widetilde{\beta}}}

\newcommand{\wbOmega}{\boldsymbol{\widetilde{\Omega}}}

\newcommand{\ben}{\begin{enumerate}}
\newcommand{\een}{\end{enumerate}}
\newcommand{\beq}{\begin{equation}}
\newcommand{\eeq}{\end{equation}}
\newcommand{\bde}{\begin{description}}
\newcommand{\ede}{\end{description}}

\newcommand{\bx}{{\bf x}}

\newcommand{\by}{{\bf y}}

\newtheoremstyle{slplain}
  {1\baselineskip\@plus.2\baselineskip\@minus.2\baselineskip}
  {.5\baselineskip\@plus.2\baselineskip\@minus.2\baselineskip}
  {\slshape}
  {}
  {\bfseries}
  {.}
  { }
  {}

\theoremstyle{slplain}

\newtheorem{theorem}{Theorem}

\newtheorem{example}[theorem]{Example}

\newtheorem{notation}[theorem]{Notation}

\newtheorem{proposition}[theorem]{Proposition}
\newtheorem{remark}[theorem]{Remark}

\numberwithin{theorem}{section}

\usepackage{booktabs,array}

\newcount\rowc

\usepackage[inline]{enumitem}
\setlist*[enumerate]{label=(\roman*)}
\begin{document}

    \def\spacingset#1{\renewcommand{\baselinestretch}%
    {#1}\small\normalsize} \spacingset{1}
    \title{\bf Evidence Estimation in Gaussian Graphical Models Using a Telescoping Block Decomposition of the Precision Matrix}
    \author{Anindya Bhadra\thanks{Address for correspondence: 150 N. University St., West Lafayette, IN 47906. Email: bhadra@purdue.edu} \\ 
    Department of Statistics, Purdue University \\ 
    \\
    Ksheera Sagar\\
    Department of Statistics, Purdue University\\
    \\
    David Rowe\\
    Department of Statistics, Purdue University\\
    \\
    Sayantan Banerjee\\
    OM \& QT Area, Indian Institute of Management Indore\\
    \\
    Jyotishka Datta \\
    Department of Statistics, Virginia Polytechnic Institute and State University}
    \date{}
    \maketitle

\begin{abstract}

Marginal likelihood, also known as \emph{model evidence}, is a fundamental quantity in Bayesian statistics. It is used for model selection using Bayes factors or for empirical Bayes tuning of prior hyper-parameters. Yet, the calculation of evidence has remained a longstanding open problem in Gaussian graphical models. Currently, the only feasible solutions that exist are for special cases such as the Wishart or G-Wishart, in moderate dimensions. We develop an approach based on a novel \emph{telescoping block decomposition} of the precision matrix that allows the estimation of evidence by application of Chib's technique under a very broad class of priors under mild requirements. Specifically, the requirements are: (a) the priors on the diagonal terms on the precision matrix can be written as gamma or scale mixtures of gamma random variables and (b) those on the off-diagonal terms can be represented as normal or scale mixtures of normal. This includes structured priors such as the Wishart or G-Wishart, and more recently introduced element-wise priors, such as the Bayesian graphical lasso and the graphical horseshoe. Among these, the true marginal is known in an analytically closed form for Wishart, providing a useful validation of our approach. For the general setting of the other three, and several more priors satisfying conditions (a) and (b) above, the calculation of evidence has remained an open question that this article resolves under a unifying framework.
\end{abstract}

    \noindent%
    \newpage
    \spacingset{1.2}
\section{Introduction}
Marginal likelihood, also known as \emph{model evidence}, is a fundamental quantity in Bayesian statistics and its calculation is important for a number of reasons~\citep{llorente2023safe}. Maximizing the marginal likelihood provides one approach for selecting prior hyper-parameters in empirical Bayes type procedures, dating back to \citet{robbins1954empirical}. But, perhaps more importantly, marginal likelihood forms the basis of Bayesian model comparison, via Bayes factors. The lack of a viable expression for evidence often necessitates a more tractable evidence lower bound (ELBO), and forms the basis of variational Bayes approaches; see \citet{blei2017variational} for a recent review. Despite this fundamental importance, the calculation of marginal likelihood in Gaussian graphical models (GGMs) is an unresolved problem, except for very specific \emph{conjugate priors} on the precision matrix belonging to the Wishart family, such as the Wishart or G-Wishart \citep{Atay05, uhler2018exact}. The chief difficulty lies with the domain of integration, the space of positive definite matrices, that is not amenable to direct integration. The main methodological contribution of this paper is the development of a novel \emph{telescoping block decomposition} of the precision matrix that allows the estimation of evidence via an application of \citetalias{chib1995marginal} method under a class of priors considerably broader than the Wishart family or its aforementioned variants. The proposed decomposition draws inspiration from the approach of \citet{wang2012bayesian}, developed in the context of posterior sampling under the non-conjugate Bayesian graphical lasso (BGL) prior. A key innovation lies in the realization that the approach of  \citet{wang2012bayesian} applies to a broad range of {\color{black}{proper}} priors apart from BGL under mild conditions, and, with appropriate modifications, it can be used for likelihood evaluation, in addition to sampling. Through some \emph{reverse engineering}, it becomes apparent that the main requirements of our approach are: (a) the priors on the diagonal terms on the precision matrix can be written as gamma or scale mixtures of gamma random variables and (b) those on the off-diagonal terms can be represented as normal or scale mixtures of normal. This includes structured priors such as the Wishart and G-Wishart, and element-wise priors such as the BGL and the graphical horseshoe \citep[GHS,][]{li2019graphical}. Among these, the marginal likelihood under the Wishart model is known in closed form. Consequently, the Wishart case provides a useful validation of the proposed approach. For BGL, GHS and several other related priors, the calculation of marginal likelihood has remained an elusive open question. This article provides a resolution under a single unifying framework.  

\subsection{Computing Evidence in GGMs: Limitations of Generic Approaches} \label{sec:refevidence} The calculation of evidence is simple in principle: for generic density $f$, parameter $\theta$, and observed data $\by$, it is given by $f(\by) = \int f(\by\mid \theta) f(\theta) d\theta$. However, the integral may be high-dimensional. {\color{black} Moreover, when $\theta$ denotes the precision matrix of a GGM, which is of key interest in this work, the domain of integration must be restricted to the space of positive definite matrices.} This makes a forward integration all but infeasible, unless the model admits special structures such as \emph{decomposability} \citep{dawid1993hyper} or under the case of \emph{G-Wishart} \citep{Atay05, uhler2018exact}. Current state-of-the-art methods for general GGMs rely on pseudolikelihood or variational schemes that do not target the true marginal; see \citet{leppa2017learning} and the references therein. Some of the main computational approaches for estimating marginal likelihood are: the harmonic mean (HM) and modified harmonic mean estimators \citep{newton1994approximate, gelfand94} with $\alpha$-stable scaling limits under mild conditions, and hence, potentially unbounded variance \citep{wolpert2012alpha}; bridge and path sampling \citep{meng1996simulating, gelman1998simulating} and their warped versions \citep{meng2002warp}; annealed importance sampling or AIS \citep{neal2001annealed}; nested sampling \citep{skilling2006nested}; and the method of \citet{chib1995marginal} and \citet{chib2001marginal} based on Markov chain Monte Carlo (MCMC) samples. More recent and comprehensive reviews are provided by \citet{friel2012estimating} and \citet{llorente2020marginal}. Another useful synthesis is by \citet{polson2014vertical}, who place bridge, path and nested sampling under a general framework of importance sampling based approaches that perform poorly if the  importance or bridge densities are not carefully chosen. This directly gets to the heart of the problem in a GGM, in that the selection of a good importance or bridge density is far from clear under a positive definite restriction on the precision matrix, to the point that we are not aware of \emph{any} general recommendations. This is because the posterior is likely highly multi-modal with other irregular features under relatively common priors. The difficulties with choosing these densities explain, at least partially, why the literature on estimating evidence in GGMs is scant, despite no dearth of generic algorithms~\citep{luca2023adaptive}, whose failures are rather conspicuous in our numerical experiments in the subsequent sections. Another case in point is the existence of a tailored Monte Carlo method for computing evidence under G-Wishart (but \emph{only} under G-Wishart) by \citet{Atay05}, appearing almost a decade after the papers on generic HM, bridge and path sampling algorithms. The method of \citet{chib1995marginal}, however, circumvents this difficult tuning of a covering bridge or importance density, since it is solely based on MCMC posterior draws. This is not to say \citetalias{chib1995marginal} approach is without blemish: its Achilles' heel is finite mixture models where it fails due to label switching \citep{neal99}. Nevertheless, for continuous mixtures of the type we consider, \citet{chib1995marginal} holds considerable appeal while generally avoiding the pitfalls of importance sampling based approaches, and is our weapon of choice for this paper. This observation echoes that of \citet{sinharay2005empirical}, who provide extensive empirical performance comparisons for various marginal likelihood computation methods for generalized linear mixed models, before concluding: \emph{``Chib's method does, however, have one advantage over
importance and bridge sampling in that it does not require that a matching or importance sampling density be selected. If the posterior distribution has features, like an unusually
long tail, not addressed by our warping transformations, then it is possible that the standard
deviation of importance and bridge sampling may be underestimated.''}

\subsection{An Overview of  \texorpdfstring{\citet{chib1995marginal}}{chib1995orig}}
Chib relies on the fundamental Bayesian identity:
\small
\begin{eqnarray*}
f(\by) = \frac{f(\by\mid \theta)f(\theta)}{f(\theta\mid \by)},
\end{eqnarray*}
\normalsize
sometimes also called ``Candidate's formula'' with the marginal displayed on the left in this manner \citep{besag1989candidate}. Assume that the likelihood and the prior can be evaluated at some user-defined $\theta=\theta^*$ (say, the posterior mean available from MCMC). If the posterior density can also be evaluated at the same $\theta^*$, then of course the calculation is trivial. But a closed form evaluation of the posterior is unavailable apart from the simplest of models. Chib's method relies on a Gibbs sampler to estimate the posterior density at the chosen $\theta^*$ and then gives the marginal via Candidate's formula. The key point here is the denominator, the posterior density, needs to be ``evaluated;'' simply designing a Gibbs sampler to generate posterior samples is not enough. This is a fundamental difference with common MCMC sampling approaches, where it is typically enough to have an un-normalized density. However, for density evaluation, the constants must be accounted for. Chib does the following: let $\theta$ be the parameter of interest and $z$ be a collection of all other latent variables. Suppose a Gibbs sampler iteratively samples from $f(z\mid \by,\theta)$ and $f(\theta\mid z,\by)$. By standard MCMC theory, under good mixing, eventually the sampler will produce draws from $f(z,\theta\mid \by)$ with correct marginals for $(z\mid \by)$ and $(\theta\mid \by)$. Then, Chib's approximation for the denominator at $\theta^*$ is the following:
\small
\vspace{-0.1cm}
\begin{eqnarray*}
\hat f(\theta^*\mid \by) = M^{-1} \sum_{i=1}^{M} f(\theta^*\mid \by, z^{(i)}),
\end{eqnarray*}
\vspace{-0.1cm}
\normalsize
where $z^{(i)}$ is a draw from $f(z\mid \by)$ that the Gibbs sampler provides. \emph{The main challenge is that} $f(\theta \mid \by,z)$ \emph{still needs to be ``evaluated'' and the success of the method depends on identifying such a conditional posterior.}  It is nontrivial to identify and overcome this hurdle in graphical models, and, in this sense, application of Chib's method is slightly a matter of art. Candidate's formula holds for \emph{any choice} of $\theta$. But for a GGM parameterized by its precision matrix $\bOmega$, the challenge lies in partitioning $\bOmega$ into the parameter of interest, $\theta$, and the nuisance parameter, $z$. However, assuming this hurdle could be overcome, the advantages of \citetalias{chib1995marginal} method are also apparent. The procedure is \emph{automatic} in the same sense a Gibbs sampler is automatic but an accept--reject sampler requiring a choice of a proposal density is not: there are no covering densities to tune, unlike in importance or bridge sampling.  Moreover, \citetalias{chib1995marginal} method can also be viewed as an interesting special case of bridge sampling with an automatically determined bridge density, a connection made explicit in Sections 4.2.1 and 4.2.2 of \citet{llorente2020marginal}. 

\subsection{Organization of the Article and Summary of Our Contributions}
{\color{black}Our main contributions can be summarized as follows.
\begin{enumerate}
    \item Construction of a novel telescoping block decomposition of the precision matrix:
    \begin{itemize}
        \item[] We begin by delineating the proposed telescoping block decomposition in Section~\ref{sec:wang}. This lies at the heart of our Chib type decomposition of $\bOmega$ into $(z,\theta)$. Specifially, we show that under a GGM, the log marginal likelihood is given by a row or column-wise telescoping sum involving four terms. The first term is an easily computable partial likelihood evaluation (a univariate normal), irrespective of the prior, under a certain Schur complement adjustment of the precision matrix closely related to the iterative proportional scaling algorithm \citep[][pp.~134--135]{lauritzen1996graphical}. The second term is problematic and there is no easy way to evaluate it. However, by construction, this is the telescoping term that is eliminated from the overall sum, without the need for ever actually having to evaluate it.
    \end{itemize}
    \item Computation of evidence under a broad class of Gaussian mixture priors:
    \begin{itemize}
        \item[] Evaluations of the third and fourth terms in the aforementioned telescoping sum are prior-specific. We show how to compute them for Wishart, two element-wise priors: BGL and GHS, and G-Wishart in Sections~\ref{sec:wishart}, \ref{sec:element}, and \ref{sec:gwishart}, respectively. We also provide numerical demonstrations and comparisons with competing approaches for each. 
        Through the expositions in Section~\ref{sec:wishart}--\ref{sec:gwishart}, it becomes clear that the requirements for our technique to hold are simply that (a) the prior on off-diagonal terms of $\bOmega$ are scale mixtures of normal and (b) the prior on diagonal terms of $\bOmega$ are scale mixtures of gamma, a priori, shedding some light on other types of priors where our method is  applicable (see the discussion in Section~\ref{sec:conc}). 
    \end{itemize}
    \item Applications:
    \begin{itemize}
        \item []Section~\ref{sec:app} details some applications of the proposed approach in Bayesian hypothesis testing, in analyzing two real data sets, and in designing a new sampler for the G-Wishart distribution. Further, in Section~\ref{sec:multivariatet-t}, we demonstrate the applicability of the proposed technique to Gaussian scale mixture likelihood (e.g., the multivariate-$t$), which broadens the scope of our procedure to include non-Gaussian likelihoods that admit a multivariate Gaussian mixture representation for the purpose of evidence computation.
    \end{itemize}
\end{enumerate}
 The Supplementary Material contains additional technical details, computational times of competing approaches and MCMC diagnostics for Chib's method.
 }
\section{A Telescoping Block Decomposition of the Precision Matrix} \label{sec:wang}
\begin{figure}[!t]
\begin{minipage}{0.4\textwidth}
    (a)\\
    \includegraphics[height=3.75cm,width=6cm]{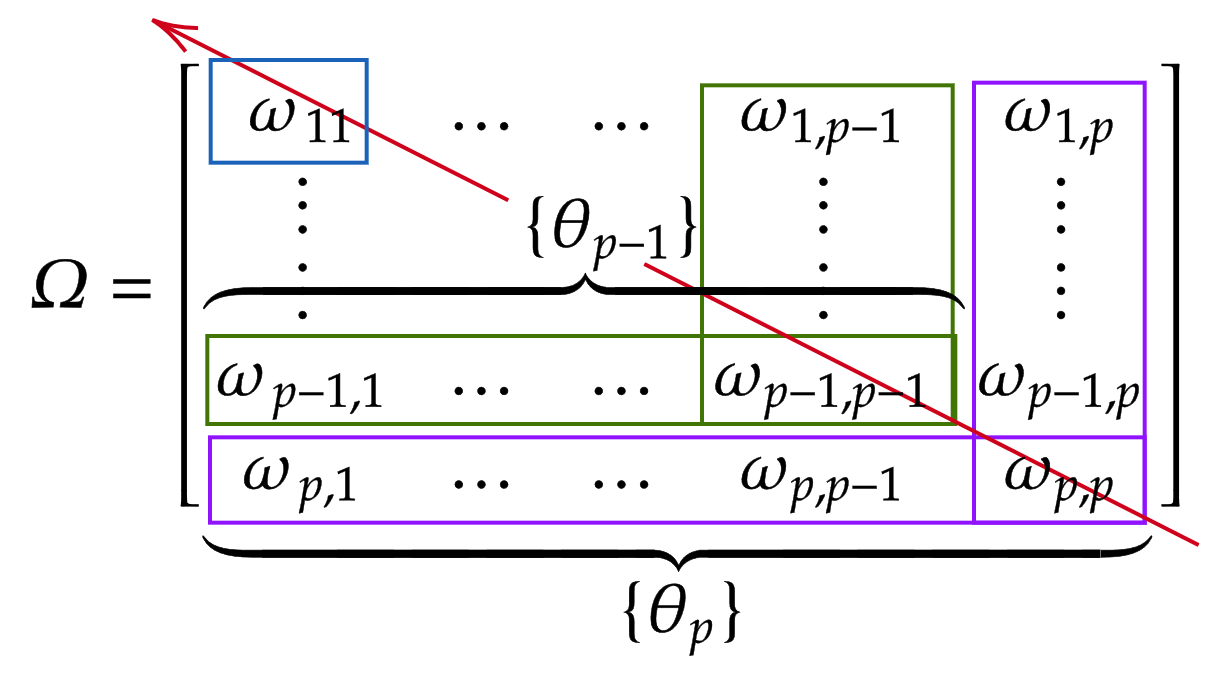}
\end{minipage}
\begin{minipage}{0.4\textwidth}
(b)\\
\tiny
\begin{equation*}
\begin{array}{ccccccccc}
        \log f(\y_{1:p}) & =& \mathrm{I}_{p}& +& \cancel{\mathrm{II}_{p\quad}}& +& \mathrm{III}_{p}& -& \mathrm{IV}_{p}\\
    \cancel{\mathrm{II}_{p\quad}} & =&   \mathrm{I}_{p-1}& +& \cancel{\mathrm{II}_{p-1}}& +& \mathrm{III}_{p-1}& -& \mathrm{IV}_{p-1}\\
     \cancel{\mathrm{II}_{p-1}} & =&   \mathrm{I}_{p-2}& +& \cancel{\mathrm{II}_{p-2}}& +& \mathrm{III}_{p-2}& -& \mathrm{IV}_{p-2}\\
     \vdots & =&  \vdots & &  \vdots & &  \vdots & &  \vdots\\
      \vdots & =&  \vdots & +&  \vdots & +&  \vdots & -&  \vdots\\
     \vdots & =&  \vdots & &  \vdots & &  \vdots & &  \vdots\\
     \cancel{\mathrm{II}_{3\text{ }\,\,\,}} & = & \mathrm{I}_{2} & +& \cancel{\mathrm{II}_{2\text{ }\,\,\,}}& +& \mathrm{III}_{2}& -& \mathrm{IV}_{2}\\
     \cancel{\mathrm{II}_{2\text{ }\,\,\,}} & = & \mathrm{I}_{1} & +& \left(\mathrm{II}_{1}=0\right)& +& \mathrm{III}_{1}& -& \mathrm{IV}_{1}\\
     \hline
      \log f(\y_{1:p}) & =& \displaystyle\sum_{j=1}^{p}\mathrm{I}_{j} & + & 0 &+&\displaystyle\sum_{j=1}^{p}\mathrm{III}_{j} & - &\displaystyle\sum_{j=1}^{p}\mathrm{IV}_{j}\\
      \hline
\end{array}
\normalsize
\end{equation*}
\end{minipage}
\caption{(a) Decomposition of $\bOmega_{p \times p}$. Purple, green and black blocks denote $\btheta_p, \btheta_{p-1}$ and finally $\btheta_1= \omega_{11}$. Red arrow denotes how the algorithm proceeds, fixing one row/column at a time, and (b) the telescoping sum giving the log-marginal $\log f(\y_{1:p})$. \label{block_matrix_explanation}}
\end{figure}

Let, $\y\sim \mathcal{N}(\mathbf{0}, \mathbf{I}_n\otimes \bOmega_{p\times p}^{-1}),$ denote an $n\times p$ matrix where each row is an i.i.d. sample from a $p$-variate normal distribution. Let $\by_i;\; i=1,\ldots,p$ denote the $i$th column of $\mathbf{y}$. We further use the notation $\by_{k:j}$ to denote the $n\times (j-k+1)$ sub-matrix of $\by$ formed by concatenating the corresponding columns for $k< j,$ and trivially $\by_{1:p}=\by$. Apply the decomposition:
\begin{center}
    \begin{tabular}{ c c }
       $\bOmega_{p\times p} = \begin{bmatrix}
\bOmega_{(p-1) \times (p-1)} & \bomega_{\sbt\,p}\\
\bomega_{\sbt\,p}^T & \omega_{pp}\\
\end{bmatrix}$. 
\end{tabular}
\end{center}
Let $\btheta_p = (\bomega_{\sbt\,p}\,,\, \omega_{pp})$ be a vector of length $p$ denoting the last column of $\bOmega_{p\times p}$ and $z$ be collection of all latent variables. Luckily, \citet{wang2012bayesian} showed for this $\btheta_p$, the conditional posterior density $f(\btheta_p \mid \by, z) = f(\bomega_{\sbt\,p}\,,\, \omega_{pp} \mid \by, z) = f(\bomega_{\sbt\,p} \mid \by, z)f(\omega_{pp} \mid \bomega_{\sbt\,p}\,,\,\by, z) $ can be evaluated as a product of normal and gamma densities under suitable priors on $\bOmega_{p\times p}$. This finding, of seemingly tenuous relevance at best to the problem at hand, is what we seek to exploit. From Bayes theorem:
\begin{eqnarray}
\log f(\by_{1:p}) &=& \log f(\by_{1:p}\mid \btheta_p) + \log f(\btheta_p) - \log f(\btheta_p \mid \by_{1:p}).\label{eq:basic}
\end{eqnarray}
The question then becomes how one should handle the integrated likelihood $f(\by_{1:p} \mid \btheta_p)$. Since $\btheta_p$ is now a sub-matrix of $\bOmega_{p\times p}$, this likelihood evaluation is certainly not the same as evaluating a multivariate normal likelihood using the full $\bOmega_{p\times p}$. A na\"ive strategy would be to draw samples from $f\left(\bOmega_{(p-1) \times (p-1)}\mid \btheta_p\right)$ and then to perform Monte Carlo evaluation of the integrated likelihood. However, this \emph{arithmetic mean} estimator of the integrated likelihood would have large variance under mild conditions, and is unlikely to be effective given the dimension of $\bOmega_{(p-1) \times (p-1)}$. We resolve this by evaluating the required densities in one row or column at a time, and proceeding backwards starting from the $p$th row, with appropriate adjustments to $\bOmega_{p\times p}$ at each step via Schur complement. Rewrite Equation (\ref{eq:basic}) as:
\begin{eqnarray}
\log f(\by_{1:p}) &=& \log f(\by_{p}\mid \by_{1:p-1}, \btheta_p) + \log f(\by_{1:p-1}\mid \btheta_p) + \log f(\btheta_p) - \log f(\btheta_p \mid \by_{1:p})  \nonumber\\
&:=& \mathrm{I}_p + \mathrm{II}_p + \mathrm{III}_p - \mathrm{IV}_p. \label{eq:basic2}
\end{eqnarray}
We deal with each term individually. First, note that the partial likelihood is:
$$
\by_{p}\mid \by_{1:p-1}, \btheta_p \sim \mathcal{N}(-\by_{1:p-1}\bomega_{\sbt\,p}\;/\omega_{pp},\; (1/\omega_{pp}){\color{black}\I_n}),
$$
which provides a convenient route to evaluating $\mathrm{I}_p$ at the chosen $\btheta_p$. We are going to assume the prior density in $\mathrm{III}_p$ can also be evaluated at $\btheta_p$ and will detail an application of \citetalias{wang2012bayesian} result for evaluating $\mathrm{IV}_p$, so that there remains the term $\mathrm{II}_p$ to deal with. At first, the development from Equation  (\ref{eq:basic}) to (\ref{eq:basic2}) does not seem very encouraging. Apparently, we have merely replaced one integrated likelihood, $f(\by_{1:p}\mid \btheta_p)$, with another, $f(\by_{1:p-1} \mid \btheta_p)$. However, the term $\mathrm{II}_p$ further admits a telescoping decomposition, as follows:
\small
\begin{align*}
\mathrm{II}_p &= \log f(\by_{p-1}\mid \by_{1:p-2}, \btheta_p, \btheta_{p-1}) + \log f(\by_{1:p-2}\mid \btheta_p, \btheta_{p-1}) + \log f(\btheta_{p-1}\mid \btheta_p) - \log f(\btheta_{p-1} \mid \by_{1:p-1}, \btheta_p)
\\
&:= \mathrm{I}_{p-1} + \mathrm{II}_{p-1} + \mathrm{III}_{p-1} - \mathrm{IV}_{p-1}.
\end{align*}
\normalsize
\sloppy
In calculating $\mathrm{I}_{p-1}$ one needs to be slightly careful since the evaluation of $ f(\by_{p-1}\mid \by_{1:p-2}, \btheta_p, \btheta_{p-1})$ is not equal to the evaluation of the univariate normal $\mathcal{N}\left(- \by_{1:p-2}\bomega_{\sbt\,(p-1)}\;/\omega_{(p-1)(p-1)},\;  (1/\omega_{(p-1)(p-1)}){\color{black}\I_n}\right)$; where $\btheta_{p-1} = (\bomega_{\sbt\,(p-1)}\; , \omega_{(p-1)(p-1)}) $ is the last column of $\bOmega_{(p-1)\times(p-1)}$. Since we are not conditioning on $\by_p$, we must not start from the node-conditional likelihood resulting from the full data, instead we need the likelihood of $f(\by_{1:p-1}\mid \btheta_p,\, \bOmega_{(p-1)\times(p-1)})$ as a starting point, a similarity shared with iterative proportional scaling \citep[][pp.~134--135]{lauritzen1996graphical}. Thus, define $\wbOmega_{(p-1)\times (p-1)}$ as:
\begin{align}
\label{Omega_tilde_p_minus_1}
     \wbOmega_{(p-1)\times (p-1)} &= \bOmega_{(p-1) \times (p-1)} -  \frac{\bomega_{\sbt\,p}\, \bomega_{\sbt\,p}^T}{\omega_{pp}}
    := \begin{bmatrix}
\wbOmega_{(p-2) \times (p-2)} & \wbomega_{\sbt\,(p-1)}\\
\wbomega_{\sbt\,(p-1)}^T & \womega_{(p-1)(p-1)}\\
\end{bmatrix},
\end{align}
so that $(\by_{1:p-1}\mid \btheta_p,\, \bOmega_{(p-1)\times(p-1)})$ is a multivariate normal with precision matrix $\wbOmega_{(p-1)\times (p-1)}$ \citep[Appendix C, ][]{lauritzen1996graphical}. Now, let the $(p-1)$ dimensional vector $\wbtheta_{p-1} = (\wbomega_{\sbt\,(p-1)}, \womega_{(p-1)(p-1)})$ be the last column of $\wbOmega_{(p-1) \times (p-1)}$. The key is to note that $\wbtheta_{p-1}$ depends only on $\btheta_p$ and $\btheta_{p-1}$ and not on the upper left $(p-2) \times (p-2)$ block of $\bOmega_{(p-1)\times(p-1)}$.
Hence, evaluation of $\mathrm{I}_{p-1}$ is possible solely as function of $\btheta_p$ and $\btheta_{p-1}$, and is independent of $\btheta_1,\ldots, \btheta_{p-2}$.  Specifically,
\begin{align*}
\by_{p-1}\mid \by_{1:p-2},\,\btheta_{p},\,\btheta_{p-1} & \sim \mathcal{N}(-\by_{1:p-2}\wbomega_{\sbt\,(p-1)}\; /\womega_{(p-1)(p-1)},\;  (1/\womega_{(p-1)(p-1)}){\color{black}\I_n}).
\end{align*}
Continuing in this manner, we evaluate $\mathrm{I}_{j}$ as a function of $\btheta_{j},\ldots,\btheta_{p}$. Calculations for the terms $\mathrm{III}$ and $\mathrm{IV}$ are prior-specific. However, we demonstrate in the next sections that it is possible to evaluate them for a wide range of commonly used priors. Thus, in each equation, only the terms $\mathrm{I},\,\mathrm{III}$ and $\mathrm{IV}$ are evaluated. The problematic term $\mathrm{II}$ is never actually evaluated. Instead, it is eliminated via a telescoping sum (Fig.~\ref{block_matrix_explanation}(b)), {\color{black}where the terms $\mathrm{II}_j$ cancel from the sum, with $\mathrm{II}_1=0$, by definition}. The algorithm proceeds backwards starting from the $p$th row or column of $\bOmega_{p\times p}\,$, fixing one row or column at a time  (Fig.~\ref{block_matrix_explanation}(a)), and making appropriate adjustments to the Schur complement.

To summarize, and to formalize our notations, we define for $j=1,\ldots,p,$ the following terms:
\small
\begin{equation*}
\setlength\arraycolsep{2pt}
    \begin{matrix*}[l]
    \mathrm{I}_{j} &=& \begin{cases} \log f(\by_{j}\mid \by_{1:j-1}, \btheta_{j},\ldots, \btheta_{p}),\; j=2,\ldots, p,\\
\log f(\by_1\mid \btheta_1,\, \btheta_2,\ldots,\btheta_p), \; j= 1,
\end{cases} &\mathrm{II}_{j} &=& \begin{cases}
\log f(\by_{1:j-1}\mid \btheta_j, \ldots, \btheta_{p}), \; j=2,\ldots, p,\\
0, \; j= 1,
\end{cases} \\
\mathrm{III}_{j} &=& \begin{cases}
\log f (\btheta_{p}), \; j=p,\\
\log f (\btheta_{j} \mid  \btheta_{j+1}, \ldots, \btheta_{p}),\; j=1,\ldots,p-1,
\end{cases}& \mathrm{IV}_{j}&=&\begin{cases}
\log f (\btheta_{p} \mid \by_{1:p}),\; j=p,\\
\log f (\btheta_{j} \mid \by_{1:j}, \btheta_{j+1}, \ldots, \btheta_{p}),\; j=1,\ldots,p-1.
\end{cases}
    \end{matrix*}
\end{equation*}
\normalsize
The desired log marginal is then given via Fig.~\ref{block_matrix_explanation}(b) as:
\begin{equation}
\label{telescoping_sum_eq}
    \log f(\y_{1:p})  = \sum_{j=1}^{p}\mathrm{I}_{j}  +  0 +\sum_{j=1}^{p}\mathrm{III}_{j}  - \sum_{j=1}^{p}\mathrm{IV}_{j}.
\end{equation}
Algorithm~\ref{algo_term_1_extra_details} specifies the details of the calculations for the first term that is procedurally independent of the prior. Throughout, $\bOmega_{j\times j}$ denotes the upper left $(j\times j)$ sub-matrix of $\bOmega$ for $j=1,\ldots,p-1$. We now proceed to demonstrate the prior-specific calculations of the third and fourth terms.
\begin{algorithm}[!t]
\caption{Computation of $\mathrm{I}_{j}$.}\label{algo_term_1_extra_details}
\begin{algorithmic}
\Require $\by_{1:p}, \bOmega_{p\times p}$.
\Ensure $\mathrm{I}_1,\ldots, \mathrm{I}_p$.
\For {(j=p,\ldots, 1)}
\If{(j==p)}
\State \Return $\mathrm{I}_{p} = \log \mathcal{N}(\by_p\mid -\by_{1:p-1}\bomega_{\sbt\,p}\;/\omega_{pp}, (1/\omega_{pp}){\color{black}\I_n}).$
\Else 
\State Set vector $\bomega_{\sbt\,(j+1)}=(\omega_{1,j+1},\ldots, \omega_{j,j+1})$. 
\State Set $\bGamma_{j \times j} = \frac{\bomega_{\sbt\,(j+1)}\,\bomega_{\sbt\,(j+1)}^{T}}{\omega_{j+1,j+1}}.$
\State Update $\bOmega_{j\times j} \leftarrow \bOmega_{j\times j} - \bGamma_{j \times j}$. 
\State Set $\wbOmega_{j\times j}=\bOmega_{j\times j}$.
\State Set $(\wbomega_{\sbt\,j}\,, \womega_{jj})$ as the last column of $\wbOmega_{j\times j}$.
\If{(j==1)}
\State \Return $\mathrm{I}_{1} = \log \mathcal{N}(\by_1\mid 0, (1/\womega_{11}){\color{black}\I_n}).$
\Else 
\State \Return $\mathrm{I}_{j} = \log \mathcal{N}\left(\by_{j} \mid -\by_{1:j-1}\wbomega_{\sbt\,j}\, /\womega_{jj},\;  (1/\womega_{jj}){\color{black}\I_n}\right).$
\EndIf
\EndIf
\EndFor
\end{algorithmic}
\end{algorithm}

\section{A Demonstration on Wishart}
\label{sec:wishart}

To avoid notational clutter, we drop the subscripts denoting dimensions when there is no scope for ambiguity (e.g., by simply using $\bOmega$), otherwise we make them explicit (e.g., by writing $\bOmega_{p\times p}$). If $\bOmega \sim \mathcal{W}_p(\mathbf{V}, \alpha)$, a Wishart prior with positive definite scale matrix $\mathbf{V}$ and degrees of freedom $\alpha>p-1$, the prior density is:
$
f(\bOmega) = \{2^{-\alpha p/2}|\V|^{-\alpha/2}/\Gamma_p(\alpha/2)\} |\bOmega|^{(\alpha-p-1)/2} \exp\{-(1/2)\mathrm{tr} (\V^{-1}\bOmega)\},
$
and the log marginal under a multivariate normal model is available in closed form as:
\begin{equation}
     \log f(\by_{1:p}) = -\frac{np}{2}\log(\pi) + \log\Gamma_p\left(\frac{\alpha+n}{2}\right) - \log\Gamma_p\left(\frac{\alpha}{2}\right) + \frac{(\alpha+n)}{2}\log\left|\I_p+\V^{1/2}\S\V^{1/2}\right|,\label{eq:truth}
 \end{equation}
where $\S =\by^T\by$,  $\Gamma_p(\cdot)$ is the multivariate gamma function and $\I_p$ is the identity matrix of size $p$. Clearly, this expression for the marginal is analytic and one does not need the proposed procedure. However, this very fact of \emph{known marginal} under Wishart also provides a useful \emph{oracle} to validate our method. We now present the details for calculating $\mathrm{III}$ and $\mathrm{IV}$ for Wishart where $\V= \mathbf{I}_p$, which is sufficient, since Wishart is a scale family. {\color{black} Specifically, if $\bOmega \sim \mathcal{W}_p(\mathbf{V}, \alpha)$ then $\mathbf{V}^{-1/2}\bOmega\mathbf{V}^{-1/2}\sim\mathcal{W}_p(\mathbf{I}_p, \alpha)$ and $\by\V^{1/2}\sim \mathcal{N}(0, (\mathbf{I}_n\kron \mathbf{V}^{-1/2}\bOmega\mathbf{V}^{-1/2})^{-1})$. One can always re-parameterize $\mathbf{W} = \mathbf{V}^{-1/2}\bOmega\mathbf{V}^{-1/2}$ and $\bx = \by\mathbf{V}^{1/2}$ to get $\bx\sim\mathcal{N}(0,\mathbf{I}_n\kron\mathbf{W}^{-1})$, where $\mathbf{W} \sim \mathcal{W}_p(\mathbf{I}_p, \alpha)$, and the log marginals under $\bx$ and $\by$ differ by an easily computed analytically available term in $\mathbf{V}$.}
\subsection{Computing \texorpdfstring{$\mathrm{III}_p$}{3p}}

Following the decomposition of $\bOmega$ and the properties of Wishart distribution \citep[Theorem 3.3.9,][]{gupta2018matrix}, if $\bOmega\sim\mathcal{W}_p(\I_p, \alpha)$ then $f(\bomega_{\sbt\,p}\,,\,\omega_{pp}) = f(\bomega_{\sbt\,p}\mid\omega_{pp})f(\omega_{pp})$, where, 
\begin{align}
\label{wishart_marginal_prior}
    \bomega_{\sbt\,p}\mid\omega_{pp} & \sim \mathcal{N}(0, \omega_{pp}\I_{p-1})\text{ and }
    \omega_{pp} \sim \mathrm{Gamma}(\text{shape = }\alpha/2, \text{rate = }1/2).
\end{align}
Thus, evaluation of $\mathrm{III}_p$ is precisely the evaluation of this product of normal and gamma densities at $\btheta_p$. We denote the chosen value of $\btheta_p$ (usually a posterior mean from MCMC) as $\btheta_p^*=(\bomega_{\sbt\,p}^*\,,\,\omega_{pp}^*)$.  
\subsection{Computing  \texorpdfstring{$\mathrm{IV}_p$}{4p}}\label{sec:wishart_IV}
Decompose $\S$ analogous to $\bOmega$ and introduce variables $\bbeta_{\sbt\,p}\,,\,\gamma_{pp}$ as: 
\begin{equation*}
    \S =\by^T\by =  \begin{bmatrix}
\S_{(p-1) \times (p-1)} & \s_{\sbt\,p}\\
\s_{\sbt\,p}^T & s_{pp}\\
\end{bmatrix},\, \bbeta_{\sbt\,p} = \bomega_{\sbt\,p}\,,\, \gamma_{pp} = \omega_{pp} - \bomega_{\sbt\,p}^T\bOmega_{(p-1)\times(p-1)}^{-1}\bomega_{\sbt\,p}.
\end{equation*}
Recall, $\btheta_p=(\bomega_{\sbt\,p}\,,\,\omega_{pp}).$ The Jacobian of transformation $(\bomega_{\sbt\,p}\,,\,\omega_{pp}) \mapsto  (\bbeta_{\sbt\,p}\,,\,\gamma_{pp})$ is 1.  Using Schur formula, $|\bOmega| = |\bOmega_{(p-1)\times(p-1)}||\omega_{pp} - \bomega_{\sbt\,p}^T\bOmega_{(p-1)\times(p-1)}^{-1}\bomega_{\sbt\,p}|$; \; $\mathrm{tr}(\S\bOmega) = 2\s_{\sbt\,p}^T\bomega_{\sbt\,p}+s_{pp}\omega_{pp}+\mathrm{tr}\Big(\S_{(p-1)\times(p-1)}\bOmega_{(p-1)\times(p-1)}\Big)$. With this, the posterior density $f(\bOmega\mid\by_{1:p})$ is: 
\small
\begin{align}
\label{beta_gamma_decomposition_general}
    f(\bOmega\mid\by_{1:p}) & \propto  f(\by_{1:p}\mid \bOmega) f(\bOmega) \propto \vert \bOmega\vert^{n/2} \exp\{-(1/2)\mathrm{tr} (\S\bOmega)\} f(\bOmega) \nonumber\\
    \propto \vert\omega_{pp} -& \bomega_{\sbt\,p}^T\bOmega_{(p-1)\times(p-1)}^{-1}\bomega_{\sbt\,p}\vert^{n/2}\left|\bOmega_{(p-1)\times(p-1)}\right|^{n/2}\exp\left(-\frac{1}{2}\Big[2\s_{\sbt\,p}^T\bomega_{\sbt\,p}+s_{pp}\omega_{pp}+\mathrm{tr}\Big(\S_{(p-1)\times(p-1)}\bOmega_{(p-1)\times(p-1)}\Big)\Big]\right)\nonumber\\
     \times \vert\omega_{pp} -& \bomega_{\sbt\,p}^T\bOmega_{(p-1)\times(p-1)}^{-1}\bomega_{\sbt\,p}\vert^{(\alpha-p-1)/2}\left|\bOmega_{(p-1)\times(p-1)}\right|^{(\alpha-p-1)/2}\exp\left(-\frac{1}{2}\Big[\omega_{pp}+\mathrm{tr}\Big(\bOmega_{(p-1)\times(p-1)}\Big)\Big]\right),
    \end{align}
    \normalsize
    where the second line in Equation (\ref{beta_gamma_decomposition_general}) shows the contribution of the likelihood to the posterior (regardless of the prior on $\bOmega$)  and the third line that of the prior (specific to Wishart). A main observation of \citet[][Section 2.4]{wang2012bayesian} is that the induced conditional posterior on $(\bbeta_{\sbt\,p},\gamma_{pp} \mid \mathrm{rest})$ has a particularly convenient form. Specifically:
    \footnotesize
    \begin{align}
    f(\bbeta_{\sbt\,p},\gamma_{pp} \mid\bOmega_{(p-1)\times(p-1)},\by_{1:p})  \propto \exp& \left(-\frac{1}{2}\Big[2\s_{\sbt\,p}^T\,\bbeta_{\sbt\,p}+(s_{pp}+1)\bbeta_{\sbt\,p}^T\bOmega_{(p-1)\times(p-1)}^{-1}\bbeta_{\sbt\,p}\Big]\right) \textstyle\gamma_{pp}^{\frac{n+\alpha-p-1}{2}}\exp\left(-\frac{1}{2}(s_{pp}+1)\gamma_{pp}\right)\nonumber\\
    = \mathcal{N}(\bbeta_{\sbt\,p}\mid -\C\s_{\sbt\,p}\,,\,\C)\times \mathrm{Gamma}&\left(\gamma_{pp} \mid \mathrm{shape = }(n+\alpha-p-1)/2+1,\mathrm{rate = } (s_{pp}+1)/{2}\right) \label{eq:betagamma},
    \end{align}
    \normalsize
where $\C =\{(s_{pp}+1)\bOmega_{(p-1)\times(p-1)}^{-1}\}^{-1}$. An important remark is in order.
\begin{remark}
\label{remark:wang}
   Equation (\ref{eq:betagamma}) can be used to \emph{sample} from the posterior of $(\bOmega_{p\times p} \mid \by_{1:p})$ via a block Gibbs sampler, by cycling over all $p$ columns. This was used by \citet{wang2012bayesian} to design a clever sampling strategy in the context of the Bayesian graphical lasso prior. Wang demonstrated that so long as the starting value is positive definite, the posterior samples of $\bOmega$ for all subsequent MCMC iterations are also positive definite. However, treating $\bOmega_{(p-1)\times(p-1)}$ as latent, a Gibbs sampler may also be utilized to \emph{evaluate} $f(\bbeta_{\sbt\,p}, \gamma_{pp} \mid \by_{1:p})$, and consequently, $f (\bomega_{\sbt\,p}\,, \omega_{pp} \mid \by_{1:p}) = f(\btheta_p\mid \by_{1:p})$, since the required normalizing constants of both normal and gamma densities are available in closed form. This is our observation. That these densities have tractable normalizing constants makes no difference to sampling following \citet{wang2012bayesian}, but is crucial for us.
 \end{remark}
\subsubsection{Approximating  \texorpdfstring{$f(\btheta_p\mid \by_{1:p})$}{c4p}}

Operationally speaking, we follow \citetalias{chib1995marginal} two block strategy. Specifically, suppose we wish to approximate $f(\btheta_p\mid \by_{1:p})=f (\bomega_{\sbt\,p}\,, \omega_{pp} \mid \by_{1:p})$ at a chosen $\btheta_p^{*} = (\bomega^*_{\sbt\,p}\,, \omega^*_{pp})$. Write $f (\bomega^*_{\sbt\,p}\,, \omega^*_{pp} \mid \by_{1:p}) = f (\bomega^*_{\sbt\,p} \mid \by_{1:p})f (\omega^*_{pp} \mid \bomega^*_{\sbt\,p}\,, \by_{1:p})$. Then, Chib's two block approximation consists of approximating the two conditional posteriors separately. First note from Equation~(\ref{eq:betagamma}) that $f (\bomega^*_{\sbt\,p} \mid \by_{1:p})$ can be approximated via MCMC samples as:
\begin{align}
\label{first_gibbs_4_gibbs}
    \hat{f}(\bomega_{\sbt\,p}^*\mid \by_{1:p}) =   \hat{f}(\bbeta_{\sbt\,p}^*\mid \by_{1:p}) & =  M^{-1}\sum_{i=1}^{M} f(\bbeta_{\sbt\,p}^*\mid \by_{1:p}, \C^{(i)}) = M^{-1}\sum_{i=1}^{M} \mathcal{N}(\bbeta_{\sbt\,p}^* \mid  -\C^{(i)} \s_{\sbt\,p}\,,\,  \C^{(i)}),
\end{align}
where $\C^{(i)}$ is the $i^\text{th}$ MCMC sample of $\C$ defined in Equation (\ref{eq:betagamma}), available via the block Gibbs strategy of \citet{wang2012bayesian}, and $\bbeta_{\sbt\,p}^*=\bomega_{\sbt\,p}^*$ is a summary statistic (we use the sample average) based on the same MCMC runs.

Next, a second, ``restricted'' version of the MCMC sampler is run with $\bomega_{\sbt\,p}$ held fixed at $\bomega^*_{\sbt\,p}$ obtained from the first, unrestricted MCMC sampler, one that was used in Equation (\ref{first_gibbs_4_gibbs}). Two subtle issues arise: first, one needs to ensure that this restricted sampler, where the non-diagonal elements in column $p$ are not updated, indeed preserves the positive definiteness of the entire $\bOmega_{p\times p}$ at every MCMC iteration. Second, one needs to estimate $f (\omega^*_{pp} \mid \bomega^*_{\sbt\,p}\,, \by_{1:p})$ using this second sampler. To address these issues, we first sample $\wbOmega_{(p-1)\times(p-1)}$ as defined in Equation~\eqref{Omega_tilde_p_minus_1}. Using this, we update $\{\bOmega_{(p-1)\times(p-1)},\; \omega_{pp}\}$ and perform the required density evaluation. The details are as follows.

Recall the definition of $\wbOmega_{(p-1)\times(p-1)}$ from Equation (\ref{Omega_tilde_p_minus_1}) and decompose $\S_{(p-1)\times(p-1)}$ as, 
\begin{equation*}
\S_{(p-1)\times(p-1)} =\begin{bmatrix}
\S_{(p-2) \times (p-2)} & \s_{\sbt\,(p-1)}\\
\s_{\sbt\,(p-1)}^T & s_{(p-1)(p-1)}\\
\end{bmatrix}.
\end{equation*}
Let $\wbbeta_{\sbt\,(p-1)} = \wbomega_{\sbt\,(p-1)}\,,\,\wgamma_{(p-1)(p-1)} =  \womega_{(p-1)(p-1)} -  \wbomega_{\sbt\,(p-1)}^T\wbOmega_{(p-2)\times(p-2)}^{-1}\wbomega_{\sbt\,(p-1)}\,$. Then the conditional posterior of $(\wbbeta_{\sbt\,(p-1)},\,\wgamma_{(p-1)(p-1)}\mid \mathrm{rest})$ can be derived analogous to Equation~\eqref{eq:betagamma} as,
\begin{align}
\label{eq:betagamma_2ndsampler}
f(\wbbeta_{\sbt\,(p-1)},\,\wgamma_{(p-1)(p-1)}\mid \wbOmega_{(p-2)\times(p-2)},\,\bomega_{\sbt\,p}^*\,,\,\omega_{pp},\,\by_{1:p})&=\quad\mathcal{N}(\wbbeta_{\sbt\,(p-1)}\mid -\widetilde{\C}\s_{\sbt\,(p-1)}\,,\,\widetilde{\C})\\
 \times  \mathrm{Gamma}(\wgamma_{(p-1)(p-1)} &\mid (n+\alpha-p-1)/2+1,\, (s_{(p-1)(p-1)}+1)/{2}),\nonumber
\end{align}
where $\widetilde{\C}=\Big\{(s_{(p-1)(p-1)}+1)\wbOmega_{(p-2)\times(p-2)}^{-1}\Big\}^{-1}$. Equation~\eqref{eq:betagamma_2ndsampler} can be used to \emph{sample} from the posterior of $(\wbOmega_{(p-1)\times(p-1)} \mid \bomega_{\sbt\,p}^{*}\,,\,\by_{1:p})$ via a block Gibbs sampler, by holding the $p$th column fixed and cycling over the remaining $(p-1)$ columns. After updating all the $(p-1)$ columns of $\wbOmega_{(p-1)\times(p-1)}$ we generate the $j$th MCMC sample from  $f(\bOmega_{(p-1)\times(p-1)},\omega_{pp} \mid \bomega_{\sbt\,p}^{*}\,,\,\by_{1:p})$ as, 
\begin{align*}
\bOmega^{(j)}_{(p-1)\times(p-1)}& \leftarrow \wbOmega^{(j)}_{(p-1)\times(p-1)} +  \bomega_{\sbt\,p}^*\,\bomega_{\sbt\,p}^{*T}/\omega_{pp}^{(j-1)}, \\
    \omega^{(j)}_{pp}\mid \bomega_{\sbt\,p}^*\,,\, \bOmega^{(j)}_{(p-1)\times(p-1)},\,\by_{1:p} &\sim   \mathrm{Gamma}\left(\frac{n+\alpha-p-1}{2}+1,\,\frac{s_{pp}+1}{2}\right) +  \bomega_{\sbt\,p}^{*T}\left(\bOmega^{(j)}_{(p-1)\times(p-1)}\right)^{-1}\bomega_{\sbt\,p}^{*}.
\end{align*}
Let $\omega_{pp}^{*}$ denote the posterior mean of $\{\omega_{pp}^{(j)}\}$ obtained from the restricted sampler that has $\bomega_{\sbt\,p}^{*}$ fixed. Then, by \citet{chib1995marginal}, $f(\omega_{pp}^*\mid \bomega_{\sbt\,p}^*,\,\by)$ can be approximated as, 
\begin{align}
\label{second_gibbs_4_gibbs}
    \hat{f}(\omega_{pp}^*\mid \bomega_{\sbt\,p}^*\,,\,\by_{1:p}) &= \frac{1}{M} \sum_{j=1}^{M}f\left(\omega_{pp}^*\mid \bomega_{\sbt\,p}^*\,,\,\by_{1:p},\,\bOmega_{(p-1)\times(p-1)}^{(j)}\right)\\
&=\frac{1}{M} \sum_{j=1}^{M}\Bigg[\mathrm{Gamma}\left(\left(\omega_{pp}^* - \bomega_{\sbt\,p}^{*T}\left(\bOmega_{(p-1)\times(p-1)}^{(j)}\right)^{-1}\bomega_{\sbt\,p}^{*}\right)\,\middle\vert\, \frac{n+\alpha-p-1}{2}+1,\,\frac{s_{pp}+1}{2}\right) \nonumber \\
& \quad \times \mathbf{1}\left\{\omega_{pp}^* > \bomega_{\sbt\,p}^{*T}\left(\bOmega_{(p-1)\times(p-1)}^{(j)}\right)^{-1}\bomega_{\sbt\,p}^*\right\}\Bigg],\nonumber
\end{align}
which can be recognized as a Monte Carlo average of truncated gamma densities, with terms not satisfying the indicator constraint contributing zero to the sum, giving a valid density evaluation. Multiplying the results of Equations (\ref{first_gibbs_4_gibbs}) and (\ref{second_gibbs_4_gibbs}) gives the desired approximation to $f(\btheta_p\mid \by_{1:p})$ at $\btheta_p=\btheta_p^{*}$.

\subsection{Computing  \texorpdfstring{$\mathrm{III}_{p-1}, \ldots, \mathrm{III}_1$}{all3p}}

Recall from Equation (\ref{Omega_tilde_p_minus_1}) that $\wbOmega_{(p-1)\times(p-1)} = \bOmega_{(p-1)\times(p-1)} - \bomega_{\sbt\,p}\,\bomega_{\sbt\,p}^T\,/\omega_{pp}\,$. We have,
\begin{align}
\label{tilda_3_wishart}
    \by_{1:(p-1)}\mid \bOmega_{(p-1)\times(p-1)},\,\bomega_{\sbt\,p}\,,\,\omega_{pp} & \sim \mathcal{N}\left(0,  \mathbf{I}_n\otimes\wbOmega_{(p-1)\times(p-1)}^{-1}\right),\nonumber\\
  \wbOmega_{(p-1)\times(p-1)} & \sim \mathcal{W}(\I_{p-1}, \alpha -1).
\end{align}
Using properties of Wishart as in Equation (\ref{wishart_marginal_prior}), 
\begin{align*}
    \wbomega_{\sbt\,(p-1)}\mid\womega_{(p-1)(p-1)} & \sim \mathcal{N}(0, \womega_{(p-1)(p-1)}\I_{p-2}),\\
    \womega_{(p-1)(p-1)} & \sim \mathrm{Gamma}(\text{shape = }(\alpha-1)/2, \text{rate = }1/2).
\end{align*}
Thus, the evaluation of $\mathrm{III}_{p-1}= f(\btheta_p\mid \btheta_{p-1})$ is simply the evaluation of this product of normal and gamma densities at $(\wbomega_{\sbt\,(p-1)}^*\,,\,\womega_{(p-1)(p-1)}^*)$, which are uniquely determined at the chosen $(\btheta_p^*,\, \btheta_{p-1}^*)$. Computations for $\mathrm{III}_{p-2}, \ldots, \mathrm{III}_1$ proceed in an identical manner and use the corresponding $\wbOmega_{(p-j)\times (p-j)}$ available from Algorithm~\ref{algo_term_1_extra_details}. 

\subsection{Computing  \texorpdfstring{$\mathrm{IV}_{p-1}, \ldots, \mathrm{IV}_{1} $}{all4p}}

On inspecting Equation \eqref{tilda_3_wishart}, it is apparent that that we have a smaller problem at hand with $p\!-\!1$ variables instead of $p$. Hence one can follow steps analogous to Equations \eqref{first_gibbs_4_gibbs} and \eqref{second_gibbs_4_gibbs} in the setting of the model in Equation \eqref{tilda_3_wishart} to compute $\mathrm{IV}_{p-1}=f(\btheta_{p-1}\mid\btheta_{p},\,\by_{1:(p-1)})$. Computations for $\mathrm{IV}_{p-2}, \ldots, \mathrm{IV}_1$ proceed in an identical manner and use the corresponding $\wbOmega_{(p-j)\times (p-j)}$ available from Algorithm~\ref{algo_term_1_extra_details}.

\subsection{Computational Complexity}
\label{sec:cplx}
Computational complexity of the proposed approach is $O(Mp^5)$, where $M$ is the number of MCMC samples. To see this, note that the dominating term in our procedure is the calculation of  $\hat f (\bomega^*_{\sbt\,p}\,, \omega^*_{pp} \mid \by_{1:p}) = \hat f (\bomega^*_{\sbt\,p} \mid \by_{1:p}) \hat f (\omega^*_{pp} \mid \bomega^*_{\sbt\,p}\,, \by_{1:p})$ according to Equations \eqref{first_gibbs_4_gibbs} and \eqref{second_gibbs_4_gibbs}. This requires evaluating a multivariate normal density, with cost $O(p^3)$; and cycling over all $p$ columns, giving a total cost of $O(p^4)$. Repeating the procedure to calculate $\hat f$ for columns $p-1,\ldots, 1$ yields the final computational cost of $O(p^5)$ per MCMC sample. Although at a first glance the computational complexity appears rather high, we provide extensive results on both statistical performances and computational times in this and subsequent sections. The statistical performance of the proposed method is the best across all the competing approaches in most settings, while being computationally scalable up to a dimension of $p=125$. Further, the computational complexity is still $O(Mp^5)$ for the subsequent sections, under priors other than Wishart, due to an analogous line of reasoning and we omit the details there.

\subsection{Numerical Experiments on Wishart}
\label{Wishart_results_sec}
All numerical results reported in this paper are based on 2.6 GHz Intel Xeon CPUs with 10 cores and 12 GB of RAM. For Wishart, we set $\V$ as a block tridiagonal matrix with entries $1/\alpha$ on the principal diagonal, $0.25/\alpha$ on the other two diagonals, and use Equation (\ref{eq:truth}) to evaluate the truth.  {\color{black}We implement the proposed procedure as described above in R, with computationally intensive parts written in C++}. Given a setting of $(p,\,n,\, \alpha)$, we draw 25 random permutations of $\{1,\ldots,p\}$. Denoting a given permutation by $\{\sigma(1),\ldots,\sigma(p)\}$, we estimate the log of marginal likelihood: $\log f(\by_{\sigma(1)},\ldots,\by_{\sigma(p)})$ for each permutation and present the mean and standard errors of our resultant log marginal estimates. For numerical experiments in this section, we choose $M=5000$ (after a burn-in of 1000 samples). This value of $M$ is chosen to be approximately the minimum number of MCMC samples that attains a pre-set upper bound of $5\times 10^{-4}$ for the absolute value of the coefficient of variation of the log marginal likelihood estimate, under all settings considered in Table~\ref{wishart_result_table}.  The implementation of the harmonic mean estimate is straightforward using the MCMC samples. Further implementation details for annealed importance sampling (AIS) and nested sampling follow. These competitors are also implemented in Sections~\ref{sec:element} and~\ref{sec:gwishart} in an identical manner, and the details are omitted there. Apart from AIS, which requires sampling from the prior, we have refrained from comparing with generic importance or bridge sampling approaches, due to a lack of obvious choices for required importance and bridge densities under a positive definite restriction. 
\begin{itemize}
    \item\textbf{AIS}: Following Equation (3) of \citet{neal2001annealed}, we construct $100$ intermediate annealed importance densities $f_t(\by)$. These are geometric averages of prior and likelihood, constructed as $f_t(\y) = f(\bOmega)\left(f(\by\mid\bOmega)\right)^t$, where $t\in\{0,0.01,\,0.02,\allowbreak\ldots,0.99,\,1\}$. The respective prior distributions, e.g., Wishart in this case, are used as proposals in all the intermediate steps. The log marginal is computed as the average of importance weights using $M$ samples. 
    \item\textbf{Nested Sampling}: We follow the approach outlined in Section 6 of \citet{skilling2006nested}, which requires sampling from progressively higher likelihood regions. We start with $M$ samples from the prior and eliminate those for which the data likelihood is below the machine precision. We perform $M$ iterations by proposing samples from the prior such that the data likelihood at the proposed sample is greater than the smallest current likelihood.   
\end{itemize}
The results in Table~\ref{wishart_result_table} indicate the proposed method has both the lowest bias and variance in all dimensions we consider. Some competitors fail to produce bounded estimates in larger dimensions.
\begin{table}[!h]
\centering
\resizebox{\textwidth}{!}{%
\begin{tabular}{cccccc}
\hline
Dimension and Parameters & Truth & Proposed & AIS  & Nested & HM 
\\ \hline
$(p =5,\,n=10,\,\alpha=7)$      & -84.13    &  -84.13 (0.04)     &  -84.3 (0.68)      &  -84.26 (0.57)         &  -82.12 (0.97)   
\\ 
$(p =10,\,n=20,\,\alpha=13)$    & -365.11   &  -365.1 (0.05)    &  -397.64  (6.1)    &  -392.2  (6.04)        &  -345.47  (1.27)   
\\ 
$(p =15,\,n=30,\,\alpha=20)$    & -837.7    &  -837.83  (0.26)    &  -1000.45  (13.5)  &  -994.87  (13.7)       &  -782.43  (0.44)   
\\ 
$(p =25,\,n=50,\,\alpha=33)$  & -2417.65 & -2416.83 (1.65) & $-\infty$ & $-\infty$ & -2235.19 (3.92)\\
$(p =30,\,n=60,\,\alpha=45)$  & -3278.93 & -3278.58 (0.93) & $-\infty$ & $-\infty$ & -3044.25 (5.17)     
\\
$(p =40,\,n=80,\,\alpha=70)$  & -5718.95 & -5718.89 (0.85) & $-\infty$ & $-\infty$ & -5328.99 (8.39)     
\\
$(p =50,\,n=75,\,\alpha=100)$  & -6422.96  &  -6422.87 (0.44) & $-\infty$ & $-\infty$ & -6012.93 (5.54) \\
\color{black} $(p =100,\,n=150,\,\alpha=200)$  & \color{black}-26046.28  &  \color{black}-26044.8 (1.98) & \color{black}$-\infty$ & \color{black}$-\infty$ & \color{black}-24270.98 (12.14) \\ 
\color{black}$(p =125,\,n=175,\,\alpha=150)$  &  \color{black}-38172.04 &  \color{black}-38169.36 (2.13) & \color{black}$-\infty$ & \color{black}$-\infty$ & \color{black}-35433.76 (17.27) 
\\
\hline
\end{tabular}
}
\caption{Mean (sd) of estimated log marginal for Wishart for the proposed approach, AIS \citep{neal2001annealed}, nested sampling \citep{skilling2006nested} and HM estimates \citep{newton1994approximate}, under 25 random permutations of the nodes $\{1,\ldots,p\}$.   Computation times for all competing procedures is given in Supplementary Table~\ref{Wishart_times_supp}.}
\label{wishart_result_table}
\end{table}
\section{Evidence under Element-wise Priors}
\label{sec:element}
    
While the demonstration on Wishart is reassuring for verifying the correctness of the proposed approach, it is also redundant for practical purposes; the marginal under Wishart is available in closed form. The natural question then is: when can a technique similar to what is presented in Section~\ref{sec:wishart} be expected to succeed in models that are hitherto intractable? A closer look at Equation (\ref{beta_gamma_decomposition_general}) reveals the answer. It is apparent that the contribution of the likelihood to the posterior does not change, regardless of what the prior is. {\color{black}However, it is the form of the likelihood that indicates what a conjugate prior is in a given parameterization.} The main advantage of the reparameterization $(\bomega_{\sbt\,j}\,,\,\omega_{jj}) \mapsto  (\bbeta_{\sbt\,j}\,,\gamma_{jj})$ is that the likelihood decomposes as (normal $\times$ gamma), for which the conjugate priors are also normal and gamma, respectively. Thus, whenever the priors on the off-diagonals $\bomega_{\sbt\,j}$ are normal (or scale mixtures of normal), and the priors on $\omega_{jj}$ are gamma (or scale mixtures of gamma) Chib's method applies in a manner near identical to Wishart, provided the corresponding latent mixing variables can be sampled. These requirements are very mild and open the door to handling a very broad class of priors. We consider two illustrative examples in this paper: the Bayesian graphical lasso or BGL \citep{wang2012bayesian} and the graphical horseshoe or GHS \citep{li2019graphical}. 
Both admit a density of the form: 
\begin{align}
\label{BGL_GHS_general_prior_structure}
    f(\bOmega\mid\btau,\,\lambda) & = C(\btau, \lambda)^{-1} \prod_{i<j}f(\omega_{ij}\mid\tau_{ij},\,\lambda)\prod_{j=1}^{p}f(\omega_{jj}\mid\lambda)\Ind{(\bOmega\in \mathcal{M}_p^+)},\nonumber\\
 f(\btau\mid\lambda) &= C(\btau,\lambda)C^{-1}\prod_{i<j}f(\tau_{ij}\mid\lambda),
\end{align}
where $\mathcal{M}_p^{+}$ denotes the space of $p\times p$ positive definite matrices and $\btau=\{\tau_{ij}\}$ is a symmetric matrix of latent mixing variables. Unlike Wishart, here the prior on $\bOmega$ is defined as a product of element-wise priors on $\omega_{ij}$ and $\omega_{jj}$, restricting the non-zero prior mass via the indicator constraint $\bOmega\in \mathcal{M}_p^+$. The appealing feature of this approach is that one can naturally encode a prior belief of sparsity in $\bOmega$ via suitable sparsity-inducing priors on the off-diagonal terms, without imposing additional structural constraints on the entire $\bOmega$. The difficulties with element-wise priors also lie in the indicator constraint, in that both sampling and likelihood evaluation become non-trivial. The former difficulty was resolved by \citet{wang2012bayesian}, as noted in Remark \ref{remark:wang}. We proceed to resolve the latter. 

Marginalizing over $\tau_{ij}$, the prior in Equation~\eqref{BGL_GHS_general_prior_structure} can be written in a more compact form as: 
\begin{equation}
\label{BGL_GHS_general_prior_structure_marginal}
    f(\bOmega\mid\lambda) = C^{-1} \prod_{i<j}f(\omega_{ij}\mid\lambda)\prod_{j=1}^{p}f(\omega_{jj}\mid\lambda)\Ind{(\bOmega\in \mathcal{M}_p^+)}.
\end{equation}
The finiteness of $C$ in Equation (\ref{BGL_GHS_general_prior_structure_marginal}) is easy to establish, so long as the priors on $\omega_{ij}$ and $\omega_{jj}$ are proper. Another crucial property is $C$ is independent of $\lambda$ whenever the priors on $\omega_{ij}$ and $\omega_{jj}$ belong to a scale family and a common scale $\lambda$ (or a constant multiple thereof)  is used for both. This can be seen via reparameterizing $(\omega_{ij},\; \omega_{jj})\mapsto (\lambda\omega_{ij},\; \lambda\omega_{jj})$, a fact also noted by \citet[][Section 2.5]{wang2012bayesian}. Thus, even if $C$ is in general intractable, it does not affect common uses of the marginal likelihood. For example, in calculation of Bayes factors, the absolute constant $C$ simply cancels from the ratio. 
For the specific case of BGL, the prior on the off-diagonal entries of $\bOmega$ is double exponential with density, $f(\omega_{ij}\mid\lambda) = (\lambda/2)\exp(-\lambda|\omega_{ij}|)$. Using a result of \citet{andrews1974scale}, the normal scale-mixture representation for this prior can be written as,
\begin{equation*}
    f_{\mathrm{BGL}}(\omega_{ij}\mid\tau_{ij},\lambda) = \frac{1}{\sqrt{2\pi\tau_{ij}}}\exp\left(-\frac{\omega_{ij}^2}{2\tau_{ij}}\right),\quad f_{\mathrm{BGL}}(\tau_{ij}\mid\lambda) = \frac{\lambda^2}{2}\exp\left(-\frac{\lambda^2}{2}\tau_{ij}\right).
\end{equation*}
In the case of GHS, the prior on the off-diagonal entries of $\bOmega$ is horseshoe. Unlike double exponential, the horseshoe prior does not have a closed form, but it still admits a normal scale mixture representation with respect to a half Cauchy mixing density~\citep{carvalho2010horseshoe}: 
\begin{equation*}
    f_{\mathrm{GHS}}(\omega_{ij}\mid\tau_{ij},\lambda) = \frac{1}{\sqrt{2\pi\tau_{ij}}}\exp\left(-\frac{\omega_{ij}^2}{2\tau_{ij}}\right)
,\quad f_{\mathrm{GHS}}(\tau_{ij}\mid\lambda) = \frac{\lambda}{\pi\sqrt{\tau_{ij}}\{1+\lambda^2\tau_{ij}\}}\Ind(\tau_{ij}>0).
\end{equation*}
For the diagonal terms, we use $f(\omega_{jj}\mid \lambda) = (\lambda/2) \exp(-\omega_{jj}\lambda/2),$ for both BGL and GHS. Both these priors enjoy excellent empirical performance in high-dimensional problems and their posterior concentration properties have recently been explored by \citet{sagar2021precision}. However, the calculation of evidence, up to an absolute multiplicative constant $C$,  has remained elusive. 
\subsection{Computing \texorpdfstring{$\mathrm{III}$}{all3p}}
Unlike Wishart, the term $\mathrm{III}=\sum_{j=1}^p \mathrm{III}_{j}$ is evaluated at the end of telescoping sum. This is because, in element-wise priors, the conditional prior densities required to evaluate $\mathrm{III}_j$ cannot be obtained in closed form, but a joint prior evaluation is easy and $\mathrm{III}$ is nothing but the evaluation of logarithm of prior density $f(\bOmega\mid\lambda)$ at $\bOmega^*$. Hence, $\sum_{j=1}^p \mathrm{III}_{j}$ in the case of BGL can be approximated as,
\begin{equation*}
    \log\hat{f}_{\mathrm{BGL}}(\bOmega^*\mid\lambda) = -\log C_{\mathrm{BGL}} + \frac{p(p-1)}{2}\log\lambda -\lambda\sum_{1\leq i<j\leq p}|\omega_{ij}^*| + p\log\frac{\lambda}{2}  - \frac{\lambda}{2}\sum_{j=1}^p \omega_{jj}^*,
\end{equation*}
whereas for GHS, the evaluation of  $f(\bOmega^*\mid\lambda)$ requires the evaluation of $f(\omega_{ij}^*\mid\lambda)$. As the horseshoe density cannot be written in a closed from, it is approximated using Monte Carlo as,
\begin{equation*}
      \hat f_{\mathrm{GHS}}(\omega_{ij}^*\mid\lambda) = \frac{1}{M}\sum_{k=1}^{M}\frac{1}{\sqrt{2\pi\tau_{ij}^{(k)}}}\exp\left(-\frac{(\omega_{ij}^*)^2}{2\tau_{ij}^{(k)}}\right),\quad\lambda\sqrt{\tau_{ij}^{(k)}} \stackrel{ind}\sim \mathcal{C}^+(0,1),
\end{equation*}
where $\mathcal{C}^+(0,1)$ is the standard half Cauchy. Hence,  $\sum_{j=1}^p \mathrm{III}_{j}$ for GHS can be approximated as,
\begin{equation*}
    \log\hat{f}_{\mathrm{GHS}}(\bOmega^*\mid\lambda) = -\log C_{\mathrm{GHS}} + \sum_{1\leq i<j\leq p}\log \hat f_{\mathrm{GHS}}(\omega_{ij}^*\mid\lambda)  + p\log \frac{\lambda}{2} -\frac{\lambda}{2}\sum_{j=1}^p \omega_{jj}^*.
\end{equation*}
\subsection{Computing \texorpdfstring{$\mathrm{IV}_p,\ldots, \mathrm{IV}_1$}{all4p}}
 Consider the case when Equation~\eqref{BGL_GHS_general_prior_structure} admits a normal scale mixture representation for the off-diagonal terms and the prior on the diagonal terms of $\bOmega$ is exponential. Thus, the computation of $\mathrm{IV}_p$ follows from~ Equations \eqref{beta_gamma_decomposition_general} and \eqref{eq:betagamma}, conditional of the latent variables $\btau=\{\tau_{ij}\}$. We have, 
\small
\begin{align}
\label{eq:betagamma_general_BGL_GHS}
f(\bbeta_{\sbt\,p},\gamma_{pp} \mid \btau_{\sbt\,p}\,,\,\bOmega_{(p-1)\times(p-1)},\by_{1:p}) 
= \mathcal{N}(\bbeta_{\sbt\,p}\mid -\C\s_{\sbt\,p}\,,\,\C)\times \mathrm{Gamma}&\left(\gamma_{pp} \mid \mathrm{shape = }n/2+1,\mathrm{rate = } (s_{pp}+\lambda)/{2}\right),
\end{align}
\normalsize
where $\C =\{\mathrm{diag}^{-1}(\btau_{\sbt\,p})+(s_{pp}+\lambda)\bOmega_{(p-1)\times(p-1)}^{-1}\}^{-1}$ and $\mathrm{diag}^{-1}(\cdot)$ represents the inverse of a diagonal matrix whose diagonal entries are  $\btau_{\sbt\,p}\,$. Thus, using Equation~\eqref{eq:betagamma_general_BGL_GHS}, we can sample $\bOmega$ from the posterior of $(\bOmega_{p\times p}\mid\by_{1:p},\,\btau)$ by cycling over all $p$ columns with the corresponding scale parameters sampled as described in \citet{wang2012bayesian} and \citet{li2019graphical} for BGL and GHS respectively. Following this,  $f(\bomega_{\sbt\,p}^*\mid \by_{1:p})$ can be approximated analogous to Equation~\eqref{first_gibbs_4_gibbs}. As in the case of Wishart, we need a second restricted sampler to approximate  $f(\omega_{pp}^*\mid\bomega_{\sbt\,p}^*,\,\by_{1:p})$. Starting with the conditional posterior of $(\wbbeta_{\sbt\,(p-1)},\,\wgamma_{(p-1)(p-1)}\mid \mathrm{rest})$, which can be derived analogous to Equation~\eqref{eq:betagamma_general_BGL_GHS}, we obtain:
\small
\begin{align}
\label{eq:betagamma_2ndsampler_BGL_GHS}
f(\wbbeta_{\sbt\,(p-1)},\,\wgamma_{(p-1)(p-1)}\mid \btau_{\sbt\,(p-1)}\,,\,\wbOmega_{(p-2)\times(p-2)},\,\bomega_{\sbt\,p}^*\,,\,\omega_{pp},\,\by_{1:p})&=\quad\mathcal{N}\left(\wbbeta_{\sbt\,(p-1)}\mid -\widetilde{\C}{\tilde\s_{\sbt\,(p-1)}}\,,\,\widetilde{\C}\right)\nonumber\\
 \times  \mathrm{Gamma}(\wgamma_{(p-1)(p-1)} &\mid n/2+1,\, (s_{(p-1)(p-1)}+\lambda)/{2}),
\end{align}
\normalsize
where $\widetilde{\C}=\Big\{\mathrm{diag}^{-1}(\btau_{\sbt\,(p-1)})+(s_{(p-1)(p-1)}+\lambda)\wbOmega_{(p-2)\times(p-2)}^{-1}\Big\}^{-1}$, ${\tilde\s_{\sbt\,(p-1)}} = \left(\s_{\sbt\,(p-1)}+\btau_{\sbt\,(p-1)}^{-1}\f_{\sbt\,(p-1)}\right)$ and the vector $\f_{\sbt\,(p-1)}$ is the $(p-1)${th} column (excluding diagonal entry) of the matrix $\bomega_{\sbt\,p}^*\,\bomega_{\sbt\,p}^{*T}/\omega_{pp}$. The inverse and product operations with respect to $\btau_{\sbt\,(p-1)}$ and $\f_{\sbt\,(p-1)}$ in calculating ${\tilde\s_{\sbt\,(p-1)}}$ are element-wise. Hence, Equation~\eqref{eq:betagamma_2ndsampler_BGL_GHS} can be used to \emph{sample} from the posterior of $(\wbOmega_{(p-1)\times(p-1)} \mid \btau_{(p-1)\times(p-1)},\,\bomega_{\sbt\,p}^{*}\,,\,\by_{1:p})$ via a block Gibbs sampler, by holding the $p$th column fixed and cycling over the remaining $(p-1)$ columns. After updating all the $(p-1)$ columns of $\wbOmega_{(p-1)\times(p-1)}$ we generate the $j$th MCMC sample from  $f(\bOmega_{(p-1)\times(p-1)},\omega_{pp} \mid \bomega_{\sbt\,p}^{*}\,,\,\by_{1:p})$ as, 
\begin{align*}
\bOmega^{(j)}_{(p-1)\times(p-1)}& \leftarrow \wbOmega^{(j)}_{(p-1)\times(p-1)} +  \bomega_{\sbt\,p}^*\,\bomega_{\sbt\,p}^{*T}/\omega_{pp}^{(j-1)}, \\
    \omega^{(j)}_{pp}\mid \bomega_{\sbt\,p}^*\,,\, \bOmega^{(j)}_{(p-1)\times(p-1)},\,\by_{1:p} &\sim   \mathrm{Gamma}\left(\frac{n}{2}+1,\,\frac{s_{pp}+\lambda}{2}\right) +  \bomega_{\sbt\,p}^{*T}\left(\bOmega^{(j)}_{(p-1)\times(p-1)}\right)^{-1}\bomega_{\sbt\,p}^{*}.
\end{align*}
With the above sampling procedure, $f(\omega_{pp}^*\mid\bomega_{\sbt\,p}^*,\,\by_{1:p})$ can be approximated analogous to Equation~\eqref{second_gibbs_4_gibbs}. Calculations of terms $\mathrm{IV}_{p-1}, \ldots, \mathrm{IV}_{1}$ are similar and once again analogous to the Wishart case, apart from the presence of the mixing variables $\btau$. A detailed description is omitted.
\subsection{Numerical Experiments on the Bayesian Graphical Lasso and Graphical Horseshoe}
\label{simu_BGL_GHS}
The marginal under element-wise priors, $f(\y\mid\lambda) = \int f(\y\mid\bOmega)f(\bOmega\mid\lambda)d\bOmega$, is not available in a closed form as for Wishart. While this makes the proposed  procedure worthwhile, its validation also becomes more challenging. Nevertheless, when $p=2$, a relatively simple expression for $f(\y\mid\lambda)$ can be obtained in a closed form for both BGL and GHS via analytic integration, which allows us to calculate the truth in order to validate our method. The results are presented in Propositions~\ref{BGL_marginal_p2_lemma} and~\ref{GHS_marginal_p2_lemma} for BGL and GHS, with respective proofs in Supplementary Sections ~\ref{proof_BGL_marginal_p2_lemma} and \ref{proof_GHS_marginal_p2_lemma}. 
\begin{proposition}
\label{BGL_marginal_p2_lemma}
When $p=2$, the marginal likelihood under the Bayesian graphical lasso prior is:
\begin{equation*}
    C_{\mathrm{BGL}}^{-1}\frac{\lambda^3 \Gamma\left(\frac{n}{2}+1\right)\Gamma\left(\frac{n+3}{2}\right)}{\pi^{n-\frac{1}{2}} \Big[(\lambda+s_{11})(\lambda+s_{22})-(\lambda - |s_{12}|)^2\Big]^{(n+3)/2}}\mathbb{E}_t\left(F(t)\right),
\end{equation*}
where, 
\begin{align*}
    t & \sim \mathrm{Gamma}\left(\mathrm{shape} = \frac{n+3}{2},\,\mathrm{rate=}\frac{(\lambda+s_{11})(\lambda+s_{22})-(\lambda - |s_{12}|)^2}{2}\right),\\
    F(t) & = \Phi\Bigg[\lambda{t^{1/2}}\left(\frac{|s_{12}|}{\lambda} - 1\right)\Bigg] + \exp\left(2\lambda|s_{12}|t\right) \Phi\Bigg[-\lambda{t^{1/2}}\left(\frac{|s_{12}|}{\lambda} + 1\right)\Bigg],
\end{align*}
with $C_{\mathrm{BGL}} = \int_0^\infty {x^{1/2}}\int_{x}^{\infty}y^{-1/2}\exp(-y)dydx\approx 0.67$.
\end{proposition}

\begin{proposition}
\label{GHS_marginal_p2_lemma}
When $p=2$, the marginal likelihood under the graphical horseshoe prior is:
\begin{equation*}
     C_{\mathrm{GHS}}^{-1}\frac{\lambda \Gamma\left(\frac{n}{2}+1\right)\Gamma\left(\frac{n}{2}+1\right)}{\pi^{n+\frac{1}{2}} \Big[(\lambda+s_{11})(\lambda+s_{22})\Big]^{\frac{n}{2}+1}}\mathbb{E}_t\left(F(t)\right),
\end{equation*}
where, 
\small
\begin{align*}
    t  \sim \mathrm{Gamma}\left(\mathrm{shape} = \frac{n}{2}+1,\,\mathrm{rate=}\frac{\lambda+s_{22}}{2}\right),\,&
    F(t) =\int_0^{\frac{t}{\lambda+s_{11}}}\exp\left(\frac{ms_{12}^2}{2}\right)m^{-1/2}\left(m+\frac{t-m(\lambda+s_{11})}{\lambda^2 t}\right)^{-1} dm.
\end{align*}
\normalsize
The constant $C_{\mathrm{GHS}}$ can be obtained via a Monte Carlo approximation as, \begin{equation*}
    C_{\mathrm{GHS}} = \mathbb{E}_{(\tau,\,m)}\left(\sqrt{\frac{m}{m+\tau^2}}\right)\approx 0.64, \text{ where }\tau\sim\mathcal{C}^+(0,1),\, m\sim\exp\left(\text{rate = }\frac{1}{2}\right)\text{ and }\tau\perp m.
\end{equation*}
\end{proposition}
For numerical illustrations, we set the true precision matrix $\bOmega_0$ by sampling from the prior. The data $\by_{1:p}$ are then generated by drawing $n$ samples from the multivariate normal $\mathcal{N}(0,\bOmega_0^{-1})$. The mean and standard error of our our resultant log marginal estimates are computed as in the case of Wishart (Section~\ref{Wishart_results_sec}) and we choose $M=5000$ (after a burn-in of 1000 samples). This $M$ is chosen to be approximately the minimum number of MCMC samples that attains a pre-set upper bound of $5\times 10^{-3}$ for the absolute value of the coefficient of variation of our estimate for the log marginal. Estimation results in the case of BGL and GHS are given in the Tables~\ref{BGL_results_compact} and~\ref{GHS_results_compact} respectively, with the true marginal likelihood presented for the $p=2$ case.  For dimensions $p>2$, our results provide marginal likelihoods up to a constant $C$ independent of $\lambda$. Once again, the proposed approach remains numerically stable in large dimensions where some of the competing methods do not yield finite estimates. The harmonic mean estimate does have reasonable \emph{sample variance} in these examples, but it is known to converge to $\alpha$-stable scaling limits under mild conditions \citep{wolpert2012alpha}, with undefined population variance. It is also well established that the harmonic mean estimate tends to overestimate the marginal likelihood due to pseudo-bias, and this bias is larger in complex models~\citep{lenk2009simulation}. A similar trend is observed in the \emph{oracle} Wishart case (Table~\ref{wishart_result_table}), and in all the simulation results in Tables~\ref{BGL_results_compact}--\ref{G_Wishart_results_compact_banded}. Though some corrections for the pseudo-bias have been suggested~\citep{lenk2009simulation, pajor2013note}, they are beyond the scope of the current work.
\begin{table}[!tbh]
\centering
\resizebox{\textwidth}{!}{%
\begin{tabular}{cccccc}
\hline
Dimension and Parameters & Truth* & Proposed & AIS  & Nested & HM 
\\ \hline
 $(p =2,\, \lambda = 0.4,\, n=4)$   &  -12.33 & -12.33 (0.00)  & -12.33 (0.005)  & -12.34 (0.03)  & -12.03 (0.43)  
 \\
 $(p =2,\, \lambda = 1,\, n=5)$   &  -18.46 &  -18.45 (0.00) &  -18.46 (0.007) & -18.46 (0.03)  & -18.15 (0.35) 
 \\
 $(p =2,\, \lambda = 2,\, n=10)$   &  -40.73 &  -40.73 (0.02) &  -40.73 (0.01)  & -40.73 (0.03)  & -40.06 (0.28)   
 \\
 $(p =5,\, \lambda = 1,\, n=10)$   &  - &  -78.00 (0.03) & -77.82 (2.48)  &  -76.41 (1.93) & -67.14 (1.32)  
 \\
 $(p =10,\, \lambda = 2,\, n=20)$  &  - &  -312.13 (0.1) & -342.72 (11.52)   &  -350.14 (7.37) &  -278.58 (1.31)
 \\
 $(p =15,\, \lambda = 3,\, n=30)$   &  - &  -796.66 (0.63) & $-\infty$  &  $-\infty$ & -693.32 (2.01)  
 \\
 $(p =25,\, \lambda = 5,\, n=50)$  &  - &  -2008.34 (0.41) &  $-\infty$ & $-\infty$  &  -1778.89 (2.89) 
 \\
 $(p =30,\, \lambda = 6,\, n=60)$  &  - &  -3070.85 (9.71) & $-\infty$  & $-\infty$  &  -2701.71 (5.72) 
 \\
   $(p =40,\, \lambda = 175,\, n=90)$  &  - &  -11540.13 (4.31) & $-\infty$  & $-\infty$  &  -10901.94 (6.38)\\ 
  $(p =50,\, \lambda = 140,\, n=130)$  &  - &  -19733.65 (9.81) & $-\infty$  & $-\infty$  &-18658.68 (7.1)  
  \\
\hline
\end{tabular}
}
\caption{Mean (sd) of estimated log marginal under the Bayesian graphical lasso prior for the proposed approach, AIS \citep{neal2001annealed}, nested sampling \citep{skilling2006nested} and HM estimates \citep{newton1994approximate}. Truth* is estimated as described in Proposition~\ref{BGL_marginal_p2_lemma}. Computation times for all competing procedures is given in Supplementary Table~\ref{BGL_times_supp}.}
\label{BGL_results_compact}
\end{table}

\begin{table}[!tbh]
\centering
\resizebox{\textwidth}{!}{%
\begin{tabular}{cccccc}
\hline
Dimension and Parameters & Truth* & Proposed & AIS  & Nested & HM 
\\ \hline
 $(p =2,\, \lambda = 0.4,\, n=4)$  & -11.16   & -11.16  (0.01)  & -11.38 (0.12)  & -11.25 (0.02)  & -10.78 (0.17)  
 \\
 $(p =2,\, \lambda = 1,\, n=5)$   & -20.11  &  -20.08 (0.02) & -20.38 (0.02)  & -20.22 (0.03)  &  -19.54 (0.35) 
 \\
 $(p =2,\, \lambda = 2,\, n=10)$    &  -47.19  &  -47.18 (0.02) & -47.83 (0.03)  & -47.49 (0.05)  &  -45.94 (0.45) 
 \\
 $(p =5,\, \lambda = 1,\, n=10)$   & -   &  -60.05 (0.22) & -54.93 (0.53)  & -55.27 (0.77)  &  -53.45 (0.42) 
 \\
 $(p =10,\, \lambda = 2,\, n=20)$  & -   & -300.85 (0.77) & -331.09 (9.06)  & -321.99 (7.37)  &  -263.76 (1.32) 
 \\
 $(p =15,\, \lambda = 3,\, n=30)$    & -   &  -672.01 (1.2) &  $-\infty$ &  -713.73 (6.36) &  -598.28 (2.53) 
 \\
 $(p =25,\, \lambda = 5,\, n=50)$   &  -  & -2228.66 (12.35)  &  $-\infty$ & $-\infty$  &  -1943.06 (3.01) 
 \\
 $(p =30,\, \lambda = 6,\, n=60)$   &  -  &  -3142.74 (7.73) & $-\infty$  &  $-\infty$ &  -2745.24 (5.43) 
 \\
   $(p =40,\, \lambda = 140,\, n=90)$   &  -  &  -11648.65 (11.65) & $-\infty$  &  $-\infty$ & -10928.54 (6.86) \\ 
  $(p =50,\, \lambda = 190,\, n=120)$   &  -  &  -19851.75 (28.46) & $-\infty$  &  $-\infty$ &  -18868.73 (10.08) 
  \\
\hline
\end{tabular}
}
\caption{Mean (sd) of estimated log marginal under the graphical horseshoe prior for the proposed approach, AIS \citep{neal2001annealed}, nested sampling \citep{skilling2006nested} and HM estimates \citep{newton1994approximate}. Truth* is estimated as described in Proposition~\ref{GHS_marginal_p2_lemma}. Computation times for all competing procedures is given in Supplementary Table~\ref{GHS_times_supp}.}
\label{GHS_results_compact}
\end{table}
\normalsize

\section{Evidence under G-Wishart Priors}
\label{sec:gwishart}
The G-Wishart family \citep{roverato02} is a general class of conjugate priors on the precision matrix $\bOmega$ for a GGM, where zero restrictions are placed according to a $p\times p$ adjacency matrix $\G=\{\sg_{ij}\}$ without requiring the graph be \emph{decomposable}, providing a useful generalization of the hyper Wishart family \citep{dawid1993hyper}. Specifically, $\sg_{ij}=0 \Leftrightarrow \omega_{ij}=0$ and $\sg_{ij}=1 \Leftrightarrow \omega_{ij}\ne 0$ for $i.j \in \{1,\ldots,p\}; i\ne j$. The prior density on $\bOmega$ under a G-Wishart prior, $\mathcal{G}\mathcal{W}_G(\V,\alpha)$, given an adjacency matrix $\G$, can be written as, 
\begin{equation}
\label{G_wishart_prior_density}
    f(\bOmega\mid\G)  = I_\G\left(\alpha,\V\right)^{-1}|\bOmega|^{\alpha}\exp\left(-\frac{1}{2}\text{tr}(\V\bOmega)\right)\Ind(\bOmega\in \mathcal{M}^{+}(\G)),
\end{equation}
where $\mathcal{M}^{+}(\G)$ denotes the cone of positive definite matrices satisfying the zero restrictions according to $\G$. Here $\alpha>0$ denotes the degrees of freedom and $\V$ is a positive definite scale matrix. As $\mathcal{G}\mathcal{W}_G(\V,\alpha)$ is a conjugate prior on $\bOmega$, the posterior density $f(\bOmega\mid\y,\,\G)$ is also G-Wishart,   $\mathcal{G}\mathcal{W}_G(\V+\S,\alpha+n/2)$. The key challenge in computing the marginal likelihood in this case, is the intractability of the normalizing constant $ I_\G(\alpha,\V)$ in Equation~\eqref{G_wishart_prior_density} when $\G$ is not decomposable. Hence, the log-marginal $\log f(\y)$, is given as a difference of two intractable normalizing constants: 
\begin{equation*}
    \log f(\y) = -\frac{np}{2}\log(2\pi) + \log  I_\G(\alpha+n/2,\V+\S) - \log  I_\G(\alpha,\V).
\end{equation*}
Further, unlike Wishart, G-Wishart is not a scale family. Hence we present our method for computing the log marginal for a general scale matrix $\V$, unlike the Wishart case where it suffices to consider $\V=\mathbf{I}_p$. Decompose $\V$ and $\G$ analogous to $\bOmega$ and $\S$ as:
\begin{center}
    \begin{tabular}{ c c }
       $\V = \begin{bmatrix}
\V_{(p-1) \times (p-1)} & \v_{\sbt\,p}\\
\v_{\sbt\,p}^T & \sv_{pp}\\
\end{bmatrix},\,\quad \G = \begin{bmatrix}
\G_{(p-1) \times (p-1)} & \g_{\sbt\,p}\\
\g_{\sbt\,p}^T & \sg_{pp}\\
\end{bmatrix} $. 
\end{tabular}
\end{center}
We further introduce the following notations, giving a simple illustration in Example~\ref{example_G_Wishart_conditions}.
\begin{notation}
   Let $nb(j),\,j\in\{1,\ldots,p\}$, denote the set of neighbors of node $j$ in the graph encoded by $\G$. More precisely, $nb(j)=\{k:\, k\neq j,\,\sg_{kj}=1\}$. Let $|nb(j)|$ denote the cardinality of $nb(j)$. Similarly, let the non-neighbors of node $j$ be denoted by $nb^c(j)$, where $nb^c(j) = \{1,\ldots,p\}\setminus nb(j)$.
   \end{notation}
   \begin{notation}
   For a symmetric matrix $\A$, denote by $\A^{nb(j)}$ or $[\A]^{nb(j)}$ the symmetric sub-matrix obtained by selecting the rows and columns of $\A$ according to the row and column indices in $nb(j)$. Similarly for a column vector $\a$, denote by $\a^{nb(j)}$ or $[\a]^{nb(j)}$ as the sub-vector obtained by selecting the rows according to the indices in $ nb(j)$.
\end{notation}

\begin{example}
\label{example_G_Wishart_conditions}
    \begin{tabular}{c}
Suppose $\G$ is \begin{tikzpicture}[baseline=(current bounding box.center)]
    \node[circle, draw, fill=black] (1) [label=south west:1] at (0,0) {};
    \node[circle, draw, fill=black] (2) [label=north west:2] at (0,1) {};
    \node[circle, draw, fill=black] (3) [label=north east:3] at (1,1) {};
    \node[circle, draw, fill=black] (4) [label=south east:4] at (1,0) {};
    \draw (1) -- (2) -- (4) -- (3) -- (1);
\end{tikzpicture}. Let,  $\A = \setlength\arraycolsep{2pt}\begin{bmatrix}a_{11} &  a_{12}& a_{13}& a_{14} \\
a_{12}& a_{22}& a_{23} & a_{24}\\
a_{13}& a_{23} & a_{33}& a_{34}\\
a_{14}& a_{24}& a_{34}& a_{44}
\end{bmatrix}$. Then, $\A^{nb(1)}=\begin{bmatrix} a_{22} & a_{23} \\ a_{23} & a_{33}\end{bmatrix}.$
    \end{tabular}    
Similarly, if $\a = (a_{14},\,a_{24},\, a_{34},\,a_{44})^T$, then $\a^{nb(1)} = (a_{24},\, a_{34})^T$.
\end{example}

Before proceeding further, we present a crucial decomposition of the indicator constraint in Equation~\eqref{G_wishart_prior_density} under Schur complement adjustments. We have  $|\bOmega| = |\omega_{pp}||\wbOmega_{(p-1)\times(p-1)}|$, and,
\begin{equation}
    \label{splitting_indicator_G_wishart}
    \Ind(\bOmega\in \mathcal{M}^{+}(\G))  = \Ind(\omega_{pp}>0)\times\Ind\Big[\bomega_{\sbt\,p}^{nb(p)}\,\neq 0\Big]\times\Ind\Big[\wbOmega_{(p-1)\times(p-1)} + \bomega_{\sbt\,p}\,\omega_{pp}^{-1}\,\bomega_{\sbt\,p}^T\,\in \mathcal{M}^+\!\left(\G_{(p-1)\times(p-1)}\right)\Big].
\end{equation}
In the above display, two further remarks are in order:
\begin{remark}\label{remark:wish1}
   The indicator $\Ind\Big[\bomega_{\sbt\,p}^{nb(p)}\,\neq 0\Big]$ is one if all entries in the vector $\bomega_{\sbt\,p}^{nb(p)}$ are non-zero. 
    \end{remark}
    \begin{remark}\label{remark:wish2}
     Here $\mathcal{M}^+\!\left(\G_{(p-1)\times(p-1)}\right)$ is the cone of positive definite matrices, restricted by $\G_{(p-1)\times(p-1)}$. This specific indicator function further imposes two conditions on $\wbOmega_{(p-1)\times(p-1)}$: (a) entries in $\wbOmega_{(p-1)\times(p-1)} + \bomega_{\sbt\,p}\,\omega_{pp}^{-1}\,\bomega_{\sbt\,p}^T\,$ corresponding to zero entries in $\G_{(p-1)\times(p-1)}\,$, are zero and (b) entries in $\wbOmega_{(p-1)\times(p-1)}$ corresponding to non-zero entries in $\G_{(p-1)\times(p-1)}\,$, are free. 
\end{remark}

\subsection{Computing  \texorpdfstring{$\mathrm{III}_p$}{3p}} \label{sec:IIIp_gwishart}
With the right hand side of Equation~\eqref{splitting_indicator_G_wishart}, the prior density in Equation~\eqref{G_wishart_prior_density} can be written as, 
\begin{align}
\label{G_WIshart_prior_with_indicator_split}
    f(\bOmega\mid\G)  & \propto  |\omega_{pp}|^\alpha\exp\left(-\sv_{pp}\omega_{pp}/2\right)\Ind(\omega_{pp}>0)\nonumber\\
    &\times\exp\left(-\frac{1}{2}\Big[\bomega_{\sbt\,p}^T\,\V_{(p-1)\times(p-1)}\,\omega_{pp}^{-1}\,\bomega_{\sbt\,p}\, + 2\v_{\sbt\,p}^T\,\bomega_{\sbt\,p}\Big]\right)\Ind\Big[\bomega_{\sbt\,p}^{nb(p)}\,\neq 0\Big]\nonumber\\
    &\times|\wbOmega_{(p-1)\times(p-1)}|^\alpha\exp\left(-\frac{1}{2}\text{tr}\Big[\V_{(p-1)\times(p-1)}\wbOmega_{(p-1)\times(p-1)}\Big]\right)\nonumber\\
    & \times\Ind\Big[\wbOmega_{(p-1)\times(p-1)} + \bomega_{\sbt\,p}\,\omega_{pp}^{-1}\,\bomega_{\sbt\,p}^T\,\in \mathcal{M}^+\!\left(\G_{(p-1)\times(p-1)}\right)\Big].
\end{align}
From the above, the conditional prior $f\left(\bomega_{\sbt\,p}^{nb(p)}\,,\,\omega_{pp}\mid\G\right)$ can be written as, 
\begin{equation}
 \label{g_wishart_III_p_normal_gamma}
    f\left(\bomega_{\sbt\,p}^{nb(p)}\,,\,\omega_{pp}\mid\G\right) = \mathcal{N}\left(\bomega_{\sbt\,p}^{nb(p)}\,\middle\vert\, -\U\v_{\sbt\,p}^{nb(p)},\,\U\right)\times\text{Gamma}\left(\omega_{pp}\mid \alpha+ |nb(p)|/2 + 1,\, \sv_{pp}/2\right),
\end{equation}
where, $\U = \left[\omega_{pp}^{-1}\V_{(p-1)\times(p-1)}^{nb(p)}\right]^{-1}$. Given $\G$, all entries in $\bomega_{\sbt\,p}^{nb^c(p)}$ are zero and their contribution to the conditional prior density is a product of point masses at zero. If $nb(p)=\phi$, then the normal density in Equation~\eqref{g_wishart_III_p_normal_gamma} evaluates to 1. It is implicit that we write $\bomega_{\sbt\,p}\,$ for the sake of completeness while using Schur formula for the prior and the posterior; whereas, both the prior and the posterior are non-degenerate only for the parameters $\bomega_{\sbt\,p}^{nb(p)}\,$, given $\G$. Thus, evaluation of $\mathrm{III}_p$ is precisely the evaluation of this product of normal and gamma densities at the chosen $\left({\bomega_{\sbt\,p}^{*}}^{\,nb(p)}\,,\,\omega_{pp}^*\right)$. 
\subsection{Computing  \texorpdfstring{$\mathrm{IV}_p$}{4p}}

Using Schur formula, $|\bOmega| = |\bOmega_{(p-1)\times(p-1)}||\omega_{pp} - \bomega_{\sbt\,p}^T\bOmega_{(p-1)\times(p-1)}^{-1}\bomega_{\sbt\,p}|$, and the decomposition of the matrices $\S,\,\V,\,\G$ defined earlier, we write the posterior in the case of G-Wishart along the lines of Equation~\eqref{beta_gamma_decomposition_general} as follows:
\small
\begin{align}
\label{beta_gamma_IV_p_G_Wishart}
    f(\bOmega\mid\y,\,\G) & \propto  f(\y\mid \bOmega,\,\G) f(\bOmega\mid\G) \propto \vert \bOmega\vert^{n/2} \exp\{-(1/2)\mathrm{tr} (\S\bOmega)\} f(\bOmega\mid\G) \nonumber\\
    \propto \vert\omega_{pp} -& \bomega_{\sbt\,p}^T\bOmega_{(p-1)\times(p-1)}^{-1}\bomega_{\sbt\,p}\vert^{n/2}\left|\bOmega_{(p-1)\times(p-1)}\right|^{n/2}\exp\left(-\frac{1}{2}\Big[2\s_{\sbt\,p}^T\bomega_{\sbt\,p}+s_{pp}\omega_{pp}+\mathrm{tr}\Big(\S_{(p-1)\times(p-1)}\bOmega_{(p-1)\times(p-1)}\Big)\Big]\right)\nonumber\\
     \times \vert\omega_{pp} -& \bomega_{\sbt\,p}^T\bOmega_{(p-1)\times(p-1)}^{-1}\bomega_{\sbt\,p}\vert^\alpha\left|\bOmega_{(p-1)\times(p-1)}\right|^\alpha\exp\left(-\frac{1}{2}\Big[2\v_{\sbt\,p}^T\bomega_{\sbt\,p}+\sv_{pp}\omega_{pp}+\mathrm{tr}\Big(\V_{(p-1)\times(p-1)}\bOmega_{(p-1)\times(p-1)}\Big)\Big]\right)\nonumber\\
     \times \Ind&\left(\bOmega_{(p-1)\times(p-1)} \in \mathcal{M}^+\!\left(\G_{(p-1)\times(p-1)}\right)\right)\times \Ind(\omega_{pp} - \bomega_{\sbt\,p}^T\bOmega_{(p-1)\times(p-1)}^{-1}\bomega_{\sbt\,p}>0).
\end{align}
\normalsize
One can again use the reparemeterization of \citet{wang2012bayesian} as in Section~\ref{sec:wishart_IV} for the Wishart case, except there is now conditioning on $\G$, similar to the calculations for $\mathrm{III}_{p}$ in Section~\ref{sec:IIIp_gwishart}. The detailed calculations are presented in Supplementary Section~\ref{sec:gwishart_IV}.

\subsection{Computing  \texorpdfstring{$\mathrm{III}_{p-1}, \ldots, \mathrm{III}_1$ }{all3p} and \texorpdfstring{$\mathrm{IV}_{p-1}, \ldots, \mathrm{IV}_1$ }{all3p}}
These follow analogously to the Wishart case, conditional on $\G$, with details in Supplementary Sections~\ref{sec:gwishart_III} and~\ref{sec:gwishart_IVp}.

\subsection{Numerical Experiments on G-Wishart}

We generate a symmetric $\G$ with upper-diagonal entries from $\text{Bernoulli}(0.5)$ and fix the scale matrix, $\V=p\I_p$. For estimating evidence under G-Wishart, there exists a customized Monte Carlo method by \citet{Atay05}, implemented via the function \texttt{gnorm()} in the \texttt{R} package \texttt{BDgraph} by \citet{mohammadi2015bdgraph}, which is considered the gold standard. We use this method with $1.2\times 10^{4}$ Monte Carlo samples. Though the function \texttt{gnorm()} does not perform a maximal clique decomposition, it is sufficiently fast (see Supplementary Table~\ref{G_Wishart_times_supp}); and is a reasonable default choice without a prior knowledge for the clique structure of the graph. 
Like in the case of Wishart  (Section~\ref{Wishart_results_sec}), to achieve an upper bound of $2\times 10^{-3}$ for the coefficient of variation of our resultant estimate, we draw $M= 10^4$ MCMC samples (after a burn-in of $2000$) at every step of the telescoping sum. We also compare with generic methods for evidence calculations from the previous sections. Mean (sd) of the resulting estimates are summarized in Table~\ref{G_Wishart_results_compact}. Our method gives results very close to \citet{Atay05}, with comparable standard errors, {\color{black} although we note that the method of \citet{Atay05} failed  to yield finite estimates at $p=100$ and $125$ under these settings.}  
Of course, the method of \citet{Atay05} is specific to G-Wishart and cannot be used, for example, in the case of element-wise priors, unlike our method. We note here although there exist theoretically exact formulas for calculating evidence under G-Wishart \citep{uhler2018exact}, we have been unable to use them in any reasonably complicated graphs, and defaulted to using \citet{Atay05} as the main competitor. 
\begin{table}[!h]
\centering
\resizebox{\textwidth}{!}{%
\begin{tabular}{cccccc}
\hline
Dimension and Parameters  & Proposed & AKM & AIS  & Nested & HM 
\\ \hline
 $(p =5,\, \alpha = 2,\, n=10)$  &  -83 (0.04)  & -82.97 (0.006)  &  -84.42 (0.13) & -83.03 (0.19)  &  -81.23 (0.8)    
 \\
 $(p =10,\, \alpha = 3,\, n=20)$  & -313.68 (0.21)   & -312.85 (0.04)  & -313.42 (2.19)  & -316.76 (1.9)  & -306.29 (1.27)    
 \\
 $(p =15,\, \alpha = 5,\, n=30)$  &  -623.78 (0.57)  &  -621.75 (0.12)  & -717.57 (3.95)  &  -640.75 (3.24) &  -605.71 (1.6)   
 \\
 $(p =25,\, \alpha = 10,\, n=50)$  &  -1725.67 (0.67)  & -1723.19 (0.87)  & $-\infty$  & $-\infty$  &   -1651.93 (3.17) 
 \\
 $(p =30,\, \alpha = 20,\, n=60)$  &   -2170.26 (0.21) &  -2167.51 (0.53)  & $-\infty$  & $-\infty$  &  -2096.1 (2.77)   
 \\
  $(p =40,\, \alpha = 25,\, n=80)$  &   -3951.91 (0.38) &  -3949.03 (1.37)  & $-\infty$  & $-\infty$  &  -3807.85 (2.63)   
 \\
  $(p =50,\, \alpha = 15,\, n=100)$  & -7858.41 (1.52)   & -7895.22 (5.73)  & $-\infty$  & $-\infty$  & -7478.16 (8.51) \\ 
{\color{black}$(p =100,\, \alpha = 50,\, n=200)$}  & {\color{black}- 27155.15 (3.01)}   & ${\color{black}-\infty}$  & {\color{black}$-\infty$}  & ${\color{black}-\infty}$  &  {\color{black}-25953.92 (22.54)}\\
${\color{black}(p =125,\, \alpha = 50,\, n=250)}$  & {\color{black}- 45817.22 (4.83)}   & ${\color{black}-\infty}$  & ${\color{black}-\infty}$  & ${\color{black}-\infty}$  &  {\color{black}-43589.67 (20.13)}\\
\hline
\end{tabular}
}
\caption{(Non-decomposable $\G$, with $\sg_{ij}\sim\mathrm{Bernoulli}(0.5)$). Mean (sd) of estimated log marginal under G-Wishart for the proposed approach, AKM \citep{Atay05}, AIS \citep{neal2001annealed}, nested sampling \citep{skilling2006nested} and HM estimates \citep{newton1994approximate}. Computation times for all competing procedures is given in Supplementary Table~\ref{G_Wishart_times_supp}. }
\label{G_Wishart_results_compact}
\end{table}

For the sake of completeness, we also provide results for decomposable $\G$, when the true marginal is known \citep[Equation (45) of][]{dawid1993hyper}, providing another oracle to validate our results. We work with a tri-diagonal $\G$ and set $\V=p\I_p$. Table~\ref{G_Wishart_results_compact_banded} summarizes the results. While results from our approach and AKM are competitive in smaller dimensions, the proposed method results in both lower bias and lower variance than AKM in larger dimensions of $p=40$ and $p=50$ for this setting. {\color{black} The AKM approach again failed  to yield finite estimates at $p=100$ and $125$ under these settings.}
\begin{table}[!h]
\centering
\resizebox{\textwidth}{!}{%
\begin{tabular}{ccccccc}
\hline
Dimension and Parameters & Truth  & Proposed & AKM & AIS  & Nested & HM \\ \hline
 $(p =5,\, \alpha = 2,\, n=10)$  & -61.15 &  -61.23 (0.00) & -61.15 (0.002)  &  -62.03 (0.31) & -61.14 (0.06)  & -60.64 (0.21) \\
 $(p =10,\, \alpha = 3,\, n=20)$  & -279.57 & -279.75 (0.03)  & -279.56 (0.02)  & -311.59 (2.23)  & -279.99 (1.01)  & -276.12 (1.12)\\
 $(p =15,\, \alpha = 5,\, n=30)$  & -715.8 & -715.89 (0.01)  & -715.86 (0.14)  & $-\infty$  &  -723.33 (2.55) & -706.35 (0.97)\\
 $(p =25,\, \alpha = 10,\, n=50)$  & -1913.25 & -1913.36 (0.01)  & -1913.54 (0.32)  & $-\infty$  & $-\infty$  & -1897.61 (1.74) \\
 $(p =30,\, \alpha = 20,\, n=60)$   &  -2334.17 &  -2334.23 (0.01) & -2334.73 (0.38)  &  $-\infty$ & $-\infty$  &-2317.71 (2.02) \\
 $(p =40,\, \alpha = 25,\, n=80)$   &  -4061.4 &  -4061.37 (0.05) & -4063.61 (1.19)  &  $-\infty$ & $-\infty$  &-4026.15 (2.59) \\
  $(p =50,\, \alpha = 15 ,\, n=100)$   & -8226.84 &  -8226.95 (0.02) & -8248.24 (2.09)  &  $-\infty$ & $-\infty$  & -8163.91 (1.67) \\
  ${\color{black}(p =100,\, \alpha = 50,\, n=200)}$ &{\color{black}-28097.74} &  {\color{black}-28097.85 (0.02)}   & ${\color{black}-\infty}$  & ${\color{black}-\infty}$  & ${\color{black}-\infty}$  &  {\color{black}-28029.51 (1.96)}\\
${\color{black}(p =125,\, \alpha = 50,\, n=250)}$  & {\color{black}-47298.44} & {\color{black}-47298.5 (0.04)}   & ${\color{black}-\infty}$  & ${\color{black}-\infty}$  & ${\color{black}-\infty}$  &  {\color{black}-47178.69 (3.23)}\\
\hline
\end{tabular}
}
\caption{(Decomposable banded tri-diagonal $\G$). Mean (sd) of estimated log marginal under G-Wishart for the proposed approach, AKM \citep{Atay05}, AIS \citep{neal2001annealed}, nested sampling \citep{skilling2006nested} and HM estimates \citep{newton1994approximate}. Computation times of the competing approaches are similar to as reported in Supplementary Table~\ref{G_Wishart_times_banded_supp}.}
\label{G_Wishart_results_compact_banded}
\end{table}

\section{Applications}\label{sec:app}
\subsection{Hyperparameter Tuning via Maximum Marginal Likelihood and Bayes Factors}
\begin{figure}[!th]
     \centering
     \begin{subfigure}[b]{0.45\textwidth}
         \centering
         \includegraphics[width=\textwidth,height=5cm]{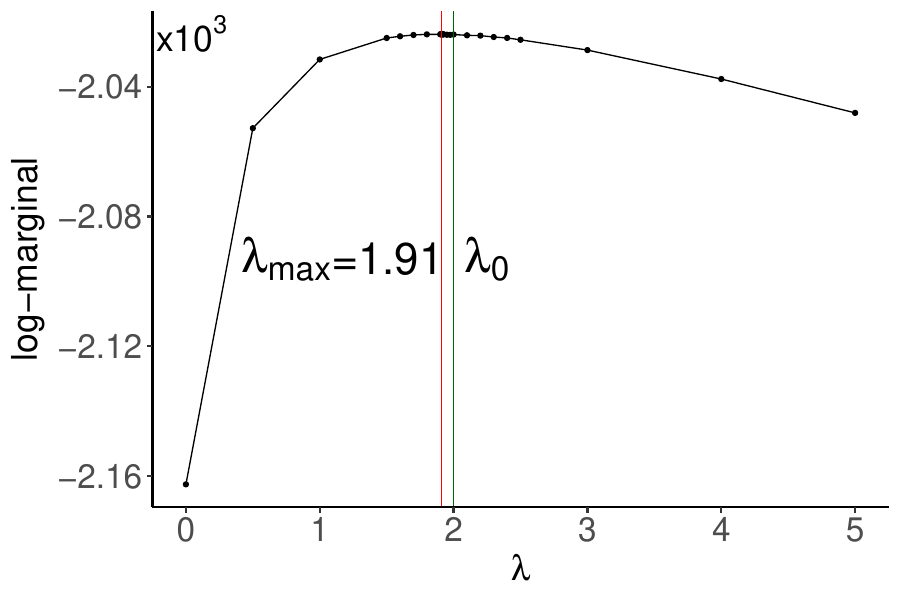}
         \caption{BGL}
         \label{BGL_MLE_vs_lambda}
     \end{subfigure}
     \hfill
     \begin{subfigure}[b]{0.45\textwidth}
         \centering
         \includegraphics[width=\textwidth,height=5cm]{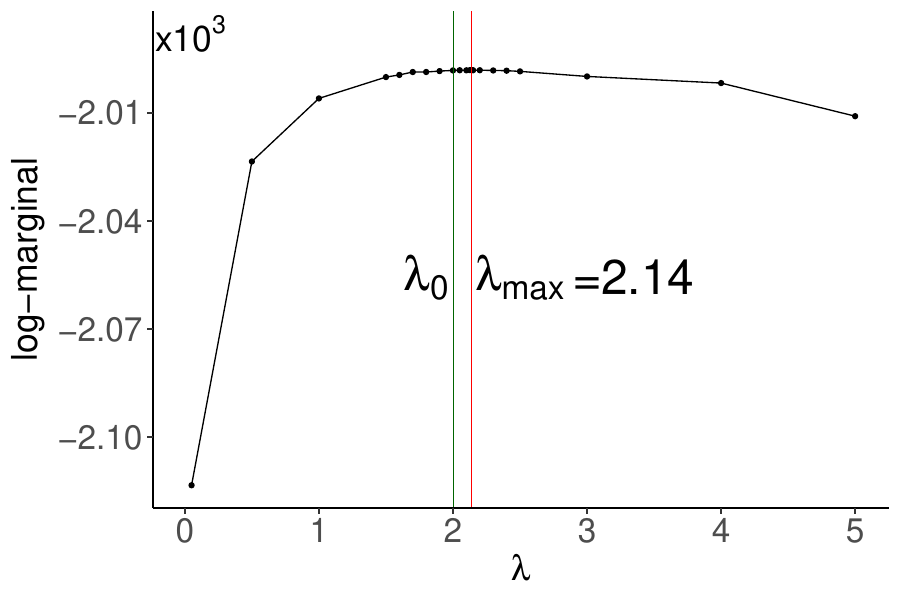}
        \caption{GHS}
         \label{GHS_MLE_vs_lambda}
     \end{subfigure}
        \caption{Log marginal likelihood vs. $\lambda$ under (a) BGL and (b) GHS.}
        \label{BGL_GHS_MLE_vs_lambda}
\end{figure}

Figure~\ref{BGL_GHS_MLE_vs_lambda} presents the log marginal likelihood estimate against the tuning parameter $\lambda$ under the BGL and GHS priors of Section~\ref{sec:element}, up to an additive absolute constant not depending on $\lambda$. Data are generated using $\lambda=\lambda_0=2$ with $p=10$ and $n=150$ under both priors.  The maximum marginal likelihood estimate (MMLE) of $\lambda$ is denoted by $\lambda_\mathrm{max}.$ We obtain for BGL, $\lambda_\mathrm{max}=1.91$, whereas for GHS, $\lambda_\mathrm{max}=2.14$. The estimates, along with the curvature of the likelihood surface, indicate this parameter is well identified. This is corroborated by the log Bayes factors presented in Table~\ref{Bayes_factor_comparison}, indicating departures from true $\lambda$ in either direction are sharply penalized. We remark that the optimality properties of the MMLE of $\lambda$ under the horseshoe prior has been studied in linear regression models \citep{van2017uncertainty}, but similar results have been hitherto unavailable for the GHS, due to a lack of a feasible algorithm for computing the MMLE in graphical models.
\begin{table}[!t]
\centering
\footnotesize
\begin{tabular}{cccccccc}
\hline
                                   & $\lambda$ & 0.05    & 1     & 2 $(=\lambda_0)$ & 3    & 4    & 5     \\\hline
\multirow{2}{*}{$\log\mathrm{BF}$} & BGL       & 138.84  & 7.86  & 0.18 & 4.98 & 13.9 & 24.34 \\
                                   & GHS       & 115.31  & 7.89  & 0.12 & 1.79 & 3.63 & 12.83 \\\hline 
\end{tabular}
\normalsize
\caption{Logarithm of Bayes factors. True $\lambda=\lambda_0=2.$}
\label{Bayes_factor_comparison}
\end{table}
\subsection{Data Applications}
We consider two applications of the proposed method on real data: the first is an inference on cell signaling network with a classical single cell flow cytometry data with a relatively modest dimension of $p=11$, and the second is an inference of protein--protein interaction network using state-of-the-art proteomics data with a larger dimension of $p=50$.
\subsubsection{Applications to Single Cell Flow Cytometry Data}
\label{sec_cytometry}
 We use the single cell flow cytometry data of \citet{sachs2005causal} on $p=11$ proteins for $n=300$ randomly chosen human immune system cells. Using this data, \citet{sachs2005causal} derived a causal cellular signaling network.  \citet{friedman2008glasso} and \citet{wang2012bayesian} used the data to infer undirected signaling networks using the frequentist and Bayesian graphical lasso respectively. {\color{black}Some other works that used the data set for graph structure learning include~\citet{peterson2015bayesian,hauser2015jointly} and~\citet{Castelletti2019IEG}. 
However, a marginal likelihood estimate under a GGM  and the framework of element-wise priors for this data set has been unavailable so far, which we present.} In order to achieve common grounds with previously developed approaches, we also present out of sample prediction results. 

We split the data into training and test sets, $\by_\mathrm{train}\text{ and }\by_\mathrm{test}$, each with $150$ data points.  Using the training set, we estimate the precision matrix $\bOmega^*$ using $(a)$ the frequentist graphical lasso implemented in the R-package `\texttt{glasso}'~\citep{R:glasso} using 5-fold cross validation $(b)$ using the MMLE estimate of $\lambda$ on $\by_\mathrm{train}$ under the Bayesian graphical lasso and considering $\bOmega^*$ at $\lambda_\mathrm{max}$ as the estimate and $(c)$ similarly using MMLE under the graphical horseshoe.

We use out of sample partial prediction loss: $\{{\sum_{j=1}^p || \by_j + \sum_{{k\in\{1,\ldots,p\}\setminus j}}\by_k\omega_{jk}^*/\omega_{jj}^*||^2}\}^{1/2}\,$, computed on the test data, and fitted likelihood on the training data, to compare the estimates, where $||\cdot||^2$ denotes the squared $\ell_2$ norm and $\bOmega^*=\{\omega_{jk}^*\}$. These results are presented in Table~\ref{Real_data_comapr}, and indicate much higher likelihood on the training set when $\lambda$ is tuned via MMLE, and similar out of sample prediction performances with the frequentist glasso tuned via cross validation, which is optimized for minimizing prediction loss \citep{efron2004estimation}.

\begin{table}[!tbh]
\centering
\footnotesize
\begin{tabular}{cccc}
\hline
    Cytometry data                   & glasso & BGL   & GHS  \\\hline
$\lambda_\mathrm{max}$ & $0.005$                        & 0.26  & 0.23 \\
Prediction loss        & 20.52                            & 20.63 & 20.5\\
$\log f(\y_\mathrm{train}\mid\lambda_\mathrm{max})$ & NA  & -949.21  & -936.91\\
$\log f(\y_\mathrm{train}\mid\bOmega^*,\,\lambda_\mathrm{max})$&  -803.09  & -769.33  &  -769.01 \\\hline
\end{tabular}
\normalsize
\caption{Comparison of prediction norm on $\by_\mathrm{test}$, using $\bOmega^*$ obtained via 5-fold cross validation for the frequentist glasso; and BGL and GHS tuned via MMLE. The log marginal likelihood at $\lambda_\mathrm{max}$ for BGL, GHS and the log of data likelihood at $(\bOmega^*,\,\lambda_\mathrm{max})$ for all methods are also given.}
\label{Real_data_comapr}
\end{table}
\subsubsection{Applications to Proteomics Data}

We use proteomics data measured using Reverse Phase Protein Array (RPPA) technology of a subset of patients with lung squamous cell carcinoma \citep{cancer2012comprehensive, cancer2014comprehensive, campbell2016distinct}, which is further streamlined and processed by \citet{ha2018personalized}. We consider the protein expression data of $p=50$ proteins of 250 randomly chosen patients, split into training and test sets, $\by_\mathrm{train}\text{ and }\by_\mathrm{test}$, each with $125$ data points.  Like in Section~\ref{sec_cytometry}, we estimate the precision matrix $\bOmega^*$ using the frequentist graphical lasso and using the MMLE estimate of $\lambda$ under the Bayesian graphical lasso and graphical horseshoe priors. We compare the resultant estimates in Table~\ref{Real_data_comapr_cancer}, in terms of out of sample prediction loss and the log likelihood of training data at $\bOmega^*$. In these results, we observe significantly higher likelihood on the training set and a smaller prediction loss, when $\lambda$ is tuned via MMLE.

\begin{table}[!tbh]
\centering
\footnotesize
\begin{tabular}{cccc}
\hline
    Proteomics data                   & glasso & BGL   & GHS  \\\hline
$\lambda_\mathrm{max}$ & 0.006                        & 0.138 & 0.109 \\
Prediction loss        & 548.91                            & 46.51 & 44.22\\
$\log f(\y_\mathrm{train}\mid\lambda_\mathrm{max})$ & NA  & -1782.31  & -1988.83\\
$\log f(\y_\mathrm{train}\mid\bOmega^*,\,\lambda_\mathrm{max})$&  -25090.7  & -930.15  &  -985.67\\\hline
\end{tabular}
\normalsize
\caption{Comparison of prediction norm on $\by_\mathrm{test}$, using $\bOmega^*$ obtained via 5-fold cross validation for the frequentist glasso; and BGL and GHS tuned via MMLE. The log marginal likelihood at $\lambda_\mathrm{max}$ for BGL, GHS and the log of data likelihood at $(\bOmega^*,\,\lambda_\mathrm{max})$ for all methods are also given.}
\label{Real_data_comapr_cancer}
\end{table}

\subsection{A Fast Column-wise Sampler for G-Wishart}
\label{sec_prior_sampler_G_wishart}
Although our main purpose in Equation~\eqref{g_wishart_III_p_normal_gamma} is prior \emph{ evaluation}, using the same equation a column-wise \emph{sampler} for G-Wishart is possible, which appears to have been unnoticed in the literature. The main advantage is a maximal clique decomposition is not required, which is known to be NP hard for a general graph. The details are presented in Algorithm \ref{prior_sampler_G_wishart}. We use $\W_{-j,-j}$ to denote the matrix obtained by removing the $j${th} row and column from the matrix $\W$. Similarly, we denote the $j${th} column of the matrices $\W$ and the scale matrix $\V$ as $\W_j$ and $\V_j$ respectively. Diagonal elements of the scale matrix $\V$ and the matrix $\W$ are denoted by $\sv_{jj}$ and $\sw_{jj}$ respectively. 
\begin{algorithm}[!h]
\caption{A column-wise sampler for $\mathcal{G}\mathcal{W}_G(\V, \alpha)$}
\label{prior_sampler_G_wishart}
\begin{algorithmic}
\Require $\V,\,\G,\,\alpha,\,M$.
\Ensure MCMC samples $\bOmega^{(1)},\ldots, \bOmega^{(M)}$.
\State Initialize $\W$ such that $\W\in \mathcal{M}^+(\G)$.
\For {(i =1,\ldots,M)}
\For {(j=1,\ldots, p)}
\State Sample $\W_j^{nb(j)}\sim\mathcal{N}\left(-\C\V_j^{nb(j)},\,\C\right)$, where $\C^{-1}=\sv_{jj} \times \Big[\W_{-j,-j}^{-1}\Big]^{nb(j)}$.
\State Sample $\gamma\sim\text{Gamma}(\alpha+1,\,\sv_{jj}/2)$.
\State Update $\sw_{jj} = \gamma + \W_j^T \W_{-j,-j}^{-1}\W_j$.
\EndFor
\State Save $\bOmega^{(i)} = \W$.
\EndFor
\end{algorithmic}
\end{algorithm}

To validate the proposed column-wise sampler, we compare the sample mean ($\bOmega^*$) obtained using Algorithm~\ref{prior_sampler_G_wishart} and the direct sampler proposed by \citet{lenkoski2013direct}, which remains one of the most popular approaches for sampling from G-Wishart. We use the same 4-cycle graph as \citet{lenkoski2013direct}, which is the smallest non-decomposable graph. Specifically, it is an undirected graph with four nodes with the edges $(1,4)$ and $(2,3)$ missing. The values of $\alpha, \V\text{ and }\G$ are as presented in Section 4.1 of \citet{lenkoski2013direct}.  We use $1\times 10^5$ MCMC samples, with a burn-in sample size of  $5\times 10^4$. \citet{lenkoski2013direct} used 10 million iterations for his direct sampler. The results are:
\begin{center}
 \resizebox{\textwidth}{!}{
    \begin{tabular}{c}
    $\bOmega^*\,(\text{Algorithm~\ref{prior_sampler_G_wishart}}) = \begin{bmatrix}
    0.7714  &  0.082 &  -0.0517    &     0\\
    0.082 &   1.1482     &    0  &  0.1506\\
   -0.0517   &      0  &  0.9042 &  -0.0857\\
         0  &  0.1506   &-0.0857  &  0.8932
    \end{bmatrix},\,\, \bOmega^*\,\text{\citep{lenkoski2013direct}} = \begin{bmatrix}
    0.7788  &  0.0826 &  -0.0516    &     0\\
    0.0826 &   1.1593     &    0  &  0.1527\\
   -0.0516   &      0  &  0.9122 &  -0.0863\\
         0  &  0.1527   &-0.0863  &  0.9024
    \end{bmatrix}.$
    \end{tabular}
    }
\end{center}
We observe the sample means are nearly identical. Similar verification is performed up to $p=50$, with $\alpha=1$, $\V = (2\alpha+\underset{j}{\text{max}}\,nb(j))\I_p$ and with upper-diagonal entries of $\G$ generated from $\text{Bernoulli}(0.5)$. The implementation in the \texttt{R} package \texttt{BDgraph} by \citet{mohammadi2015bdgraph} is used to sample $\bOmega$ according to \citet{lenkoski2013direct}. {\color{black} The direct sampler of~\citet{lenkoski2013direct} includes a tolerance parameter, which at low values demonstrates faster speed, but quite often leads to non-zero entries in samples of $\bOmega$ even though the corresponding entry in $\G$ is zero. Though such entries can be truncated to 0 based on the input tolerance, our proposed procedure requires no such post-hoc adjustment. The computational times to generate $1\times 10^5$ samples in both the methods and the Frobenius norm of sample mean differences are given in Table~\ref{compar_prior_samplers}. It can be seen that the proposed sampler for G-Wishart in Algorithm~\ref{prior_sampler_G_wishart} is $\sim 7-8$x faster than the direct sampler at a tolerance of $10^{-8}$ and is $\sim 3$x faster at a tolerance of $10^{-3}$ at comparable statistical accuracy. Similar comparisons with banded tri-diagonal $\G$ is in Table~\ref{compar_prior_samplers2}, where the proposed sampler is $\sim 13-15$x faster than the direct sampler at a tolerance of $10^{-8}$ and is $\sim 5$x faster at a tolerance of $10^{-3}$.}

\begin{table}[!h]
\color{black}
\centering
\footnotesize
\begin{tabular}{c|ccc|cc}
\hline
          &  \multicolumn{3}{c|}{Time (s)} &  \multicolumn{2}{c} {$||\bOmega^*_{\text{Alg.\,}2} -\bOmega^*_\mathrm{\text{Direct}} ||_{F}$} \\
Dimension & \begin{tabular}{c}
 Algorithm~2 \end{tabular} & $\text{Direct (tol = }10^{-8})$             & $\text{Direct (tol = }10^{-3})$             & $\text{(tol = }10^{-8})$           & $\text{(tol = }10^{-3})$\\
\hline
$p=5$ & 0.46& 2.64& 2.12& 0.02& 0.02\\
$p=10$  & 1.78& 8.13& 4.15& 0.03& 0.03\\
$p=20$ & 6.15& 40.47& 17.06& 0.12& 0.12\\
$p=30$ & 16.05& 120.07& 41.00& 0.21& 0.21\\
$p=50$ & 51.22& 465.93& 151.82& 0.43& 0.43  \\
          \hline
\end{tabular}
\normalsize
\captionsetup{labelfont={color=black}, textfont={color=black}}
\caption{Computational times to generate $10^5$ samples of $\bOmega$ using Algorithm~2 and the direct sampler~\citep{lenkoski2013direct} with tolerances$\,\in\{10^{-8},10^{-3}\}$, implemented as \texttt{rgwish()} in the R package \texttt{BDgraph}~\citep{mohammadi2015bdgraph}. Parameter settings: $\alpha=1$, $\mathbf{V} = (2\alpha+\underset{j}{\text{max}}\,nb(j))\mathbf{I}_p$ and upper-diagonal entries of $\mathbf{G}$ generated from $\text{Bernoulli}(0.5)$. Frobenius norm of the difference of sample means against Lenkoski is also presented. 
}
\label{compar_prior_samplers}
\label{compar_prior_samplers}
\end{table}

\begin{table}[!h]
\color{black}
\centering
\footnotesize
\begin{tabular}{c|ccc|cc}
\hline
          &  \multicolumn{3}{c|}{Time (s)} &  \multicolumn{2}{c} {$||\bOmega^*_{\text{Alg.\,}2} -\bOmega^*_\mathrm{\text{Direct}}||_{F}$} \\
Dimension & \begin{tabular}{c}
 Algorithm~2 \end{tabular} & $\text{Direct (tol = }10^{-8})$             & $\text{Direct (tol = }10^{-3})$             & $\text{(tol = }10^{-8})$           & $\text{(tol = }10^{-3})$\\
\hline
$p=5$ & 0.4& 2.61& 2.11& 0.01& 0.01\\
$p=10$  & 1.36& 8.45& 4.55& 0.02& 0.01\\
$p=20$ & 3.7& 34.61& 14.19& 0.03& 0.03\\
$p=30$ & 7.51& 90.98& 34.63& 0.03& 0.03\\
$p=50$ & 24.58& 354.71& 111.25& 0.05& 0.05  \\
          \hline
\end{tabular}
\normalsize
\captionsetup{labelfont={color=black}, textfont={color=black}}
\caption{Computational times to generate $10^5$ samples of $\bOmega$ using the proposed Algorithm~2 and the direct sampler~\citep{lenkoski2013direct} with tolerances$\,\in\{10^{-8},10^{-3}\}$, implemented as \texttt{rgwish()} in the R package \texttt{BDgraph}~\citep{mohammadi2015bdgraph}. Parameter settings: $\alpha=1$, $\mathbf{V} = (2\alpha+\underset{j}{\text{max}}\,nb(j))\mathbf{I}_p$ and banded tri-diagonal $\mathbf{G}$. Frobenius norm of the difference of sample means against Lenkoski is also presented.
}
\label{compar_prior_samplers2}
\end{table}
{\color{black}\subsection{Applications to Non-Gaussian Likelihoods that Admit a Gaussian Scale Mixture Representation}
\label{sec:multivariatet-t}
In this section we demonstrate another promising application of the procedure in computing evidence under \emph{non-Gaussian} likelihoods that admit a multivariate Gaussian mixture representation. As an example, consider the multivariate $t$ distribution with $\nu$ degrees of freedom, which has a well known representation as a mixture of multivariate Gaussian, in that $\by\sim t_\nu(0, \mathbf{I}_n\otimes\bOmega^{-1})$ is equivalent to $\by \mid \tau \sim \mathcal{N} (0, \tau^{-1}\mathbf{I}_n\otimes\bOmega^{-1}),\; \tau\sim \mathrm{Gamma} (\nu/2,\nu/2)$. Standard calculations show the full conditional posterior of $(\tau\mid \bOmega,\, \by)$ in this model is also gamma distributed, and hence $\tau$ can be easily sampled in the posterior via a Gibbs step. Thus, absorbing the latent $\tau$ in the likelihood to the set of already existing mixing variables in our Chib-type procedure makes it possible to compute the evidence under this model. We demonstrate the validity by computing the log marginal under the multivariate-$t$ likelihood with Bayesian graphical lasso and graphical horseshoe priors on the scale matrix $\bOmega$ in Table~\ref{multivariate_t_results_compact}. It can be seen from Table~\ref{multivariate_t_results_compact} that at high values of $\nu$, the estimates are close to those in Tables~\ref{BGL_results_compact} and~\ref{GHS_results_compact} under GGM, although differences exist from GGM estimates at low $\nu$. This is reasonable as the multivariate $t$-distribution converges to a multivariate Gaussian at large $\nu$, but displays substantially different behavior (e.g., polynomial tails) at low degrees of freedom.

We conclude this section by noting although we choose a multivariate-$t$ model for demonstration purposes, there is nothing specific to a multivariate-$t$ in the procedure, so long as a mixture representation with respect to the multivariate Gaussian exists, and the mixing variable is easy to sample in the posterior. Similar to Gaussian scale mixture priors, the class of non-Gaussian likelihoods that can be represented as a mixture of multivariate Gaussian likelihood is very broad. Some specific examples include the nonparanormal~\citep{liu2009nonparanormal}, the models considered by~\cite{bhadra2018inferring}, the Dirichlet and alternative multivariate-$t$~\citep{finegold2011, finegold2014}, among others.

\begin{table}[!h]
\color{black}
\centering
\footnotesize
\begin{tabular}{ccccc}
\hline
 & \multicolumn{3}{c}{BGL} \\
                         Dimension and Parameters        &   $\nu=5$ & $\nu=10$ & $\nu=10^4$ & Table~\ref{BGL_results_compact}\\
                 \hline
  $(p =2,\, \lambda = 1,\, n=5)$ &   -16.7 (0.02)     &     -16.97 (0.01)   &  -18.4 (0.004) &  -18.45  \\
  $(p =5,\, \lambda = 1,\, n=10)$ &   -112.38 (0.17)     &    -98.84 (0.2)    & -78.5 (0.02) &  -78.00\\
  $(p =15,\, \lambda = 3,\, n=30)$ &   -630.93 (1.95)     &    -669.11 (1.55)    &  -790.98 (0.65)     & -796.66 \\
  $(p =30,\, \lambda = 6,\, n=60)$ &    -3273.7 (9.69)    &  -3179.9 (11.02)      & 3069.5 (11.19)       &-3070.85\\
\hline
 & \multicolumn{3}{c}{GHS} \\
                 Dimension and Parameters               &
                                     $\nu=5$ & $\nu=10$ & $\nu=10^4$ & Table~\ref{GHS_results_compact}\\
                                     \hline
  $(p =2,\, \lambda = 1,\, n=5)$ &   -29.89 (0.04)     &  -26.13 (0.02)      &    -20.25 (0.01)   &-20.08\\
  $(p =5,\, \lambda = 1,\, n=10)$ &    -58.86 (0.5)      &   -58.41 (0.42)      & -59.88 (0.18)      &-60.05\\
  $(p =15,\, \lambda = 3,\, n=30)$  &  -866.45 (1.6)     &   -791.67 (1.76)     &  -675.3 (0.98)     &-672.01\\
  $(p =30,\, \lambda = 6,\, n=60)$ &  -3775.01 (12.49)      &  -3523.6 (14.11)     &     -3153.48 (11.64)  &-3142.74\\
 \hline
\end{tabular}
\normalsize
\captionsetup{labelfont={color=black}, textfont={color=black}}
\caption{Mean (sd) of estimated log marginal under the multivariate-$t$ likelihood with the priors Bayesian graphical lasso (BGL) and Graphical horseshoe (GHS) for the proposed approach. The columns marked Tables 2 and 3 give the corresponding estimates under GGM. Computation times for the proposed procedure is similar to the times presented in Supplementary Table~\ref{BGL_times_supp} and Table~\ref{GHS_times_supp} for BGL and GHS priors respectively.}
\label{multivariate_t_results_compact}
\end{table}

}
\section{Concluding Remarks}
\label{sec:conc}
Our main contribution in this paper is a general procedure based on a novel \emph{telescoping block decomposition} of the precision matrix for computing the marginal likelihood under a fairly wide variety of priors  in a GGM. The algorithm, being based on Chib's procedure, is automatic in the sense that it does not require an explicit choice for an importance density, which is notoriously hard to design for GGMs. Empirically, our approach provides numerically stable results in fairly large dimensions, in contrast to importance sampling based and harmonic mean approaches that become unstable or badly biased in large dimensions, sometimes failing to yield finite estimates. Some other competitors, such as bridge sampling, are unavailable since there are no obvious approaches for choosing the required bridge densities under a positive definite restriction. Thus, our procedure opens the door to using marginal likelihood for model comparison and tuning parameter selection purposes, a problem that has been hitherto considered intractable for GGMs apart from under very specific priors. As we pointed out earlier, the requirements are mild: the off-diagonal terms in $\bOmega$ are scale mixtures of normal and the diagonal terms are scale mixtures of gamma. This includes several priors not considered explicitly in our work. Consider for example a slight modification of the prior of \citet{wang2015scaling}, given by:
\begin{equation*}
    f(\bOmega) = C^{-1}\underset{1\leq i<j\leq p}{\prod}\left\{(1-\pi)\mathcal{N}(\omega_{ij}\mid 0,\, a\lambda)+\pi\mathcal{N}(\omega_{ij}\mid 0,\, b\lambda)\right\}\prod_{i=1}^p\text{Exp}(\omega_{ii}\mid\lambda)\Ind(\bOmega\in \mathcal{M}^+_p),
\end{equation*}
for known constants $a$ and $b$ such that $a$ is very close to zero and $b\gg a$.
The main feature of this prior is a two component discrete mixture, or the so called spike-and-slab prior, on the off-diagonal terms to encourage sparsity in $\bOmega$. However, this poses no special difficulty for our framework so long as the corresponding latent mixing variables can be sampled, following \citet{wang2015scaling}. Other priors involving a two component discrete mixture include \citet{gan2019bayesian, gan2022graphical} and \citet{shen2022posterior}. Application of the technique developed in the current paper appears feasible in all these instances and should be considered future work. {\color{black}Similarly, although in Section~\ref{sec:multivariatet-t} we demonstrate the use of the methodology for a specific non-Gaussian likelihood that admits a Gaussian mixture representation (the multivariate-$t$), application to models such as the nonparanormal \citep{liu2009nonparanormal} remains to be explored. Moreover, combining these approaches suggests new possibilities in evidence computation with very flexible Gaussian mixture priors in conjunction with very flexible Gaussian mixture likelihood functions.}

Although the main focus of this paper is on the calculation of evidence, as a consequence of our calculations of the term $\mathrm{III}$ for G-Wishart, we have also designed a new sampler for this distribution, as described in Section~\ref{sec_prior_sampler_G_wishart}. Notable previous works in this direction include \citet{lenkoski2013direct}, who requires a maximal clique decomposition, and \citet{wang2012efficient}, who provide an edge-wise sampler. Maximal clique decomposition has a worst case computational complexity that is known to be NP-hard for a general graph, but tends to work well when there are few but large cliques. On the other extreme, \citet{wang2012efficient}'s sampler requires no clique decomposition and is expected to work well when there are several isolated nodes or a large number of small cliques. Our algorithm lies somewhere in between: it is neither clique-wise, nor edge-wise. Instead, it is more aptly termed column-wise. Thus, we conjecture our method should be roughly agnostic to the connectivity of the graph and its complexity should scale mainly as a function of the dimension $p$. However, detailed investigation of this conjecture and comparing the relative pros and cons with the approaches of \citet{lenkoski2013direct}, \citet{wang2012efficient} or more recent developments such as \citet{boom21} are beyond the scope of the current, densely packed article focusing on \emph{model evidence}.
\section*{Acknowledgments}
The research of Bhadra is supported by US National Science Foundation Grant DMS-2014371.

\section*{Supplementary Material}
 The Supplementary Material contains proofs, additional technical details and numerical results. Computer code with complete documentation is publicly available via \texttt{github} at: \href{https://github.com/dp-rho/graphicalEvidence}{https://github.com/dp-rho/graphicalEvidence} 
 
\bibliographystyle{biom}
\bibliography{hs-review,ref}

\clearpage\pagebreak\newpage
\setcounter{equation}{0}
\setcounter{page}{1}
\setcounter{table}{0}
\setcounter{section}{0}
\setcounter{subsection}{0}
\setcounter{figure}{0}
\renewcommand{\theequation}{S.\arabic{equation}}
\renewcommand{\thesection}{S.\arabic{section}}
\renewcommand{\thesubsection}{S.\arabic{subsection}}
\renewcommand{\thepage}{S.\arabic{page}}
\renewcommand{\thetable}{S.\arabic{table}}
\renewcommand{\thefigure}{S.\arabic{figure}}
\begin{center}
{\LARGE{\bf Supplementary Material to\\ {\it {\color{black}Evidence Estimation in Gaussian Graphical Models Using a Telescoping Block Decomposition of the Precision Matrix}}}}
\end{center}

\section{Proof of Proposition~\ref{BGL_marginal_p2_lemma}}
\label{proof_BGL_marginal_p2_lemma}
 For $p=2$, the only latent parameter is $\tau_{12}$ corresponding to $\omega_{12}$. We denote this by $\tau$ for the sake of brevity. The domain of integration such that $\bOmega\in \mathcal{M}_2^{+}$ is set in Equation~\eqref{range_omega_ijs_2by2} as:
\begin{equation}
\label{range_omega_ijs_2by2}
    \bOmega = \begin{bmatrix}\omega_{11}&\omega_{12}\\ \omega_{12}& \omega_{22}\end{bmatrix}, \text{ such that }\omega_{12}\in\mathbb{R},\, \omega_{22}\in\mathbb{R}^+\,,\,\omega_{11}\in\left(\frac{\omega_{12}^2}{\omega_{22}}, \infty\right),\,\tau\in\mathbb{R}^+.
\end{equation}
The marginal is, $f(\y\mid\lambda) = \int f(\y\mid\bOmega)f(\bOmega\mid\tau,\,\lambda)f(\tau\mid\lambda)\,d\tau\,d\bOmega$, where,
\begin{align*}
    f(\y\mid\bOmega) & = \frac{\left(\omega_{11}\omega_{22} - \omega_{12}^2\right)^\frac{n}{2}}{(2\pi)^n}\exp\left(-\frac{1}{2}(s_{11}\omega_{11}+2s_{12}\omega_{12} + s_{22}\omega_{22})\right),\\
    f(\bOmega\mid \tau,\,\lambda)f(\tau\mid\lambda) & = \frac{\cbglinv}{\sqrt{2\pi\tau}}\exp\left(-\frac{\omega_{12}^2}{2\tau}\right)\left(\frac{\lambda}{2}\right)^2\exp\left(-\lambda\frac{(\omega_{11}+\omega_{22})}{2}\right) \frac{\lambda^2}{2}\exp\left(-\frac{\lambda^2\tau}{2}\right).
\end{align*}
Define $\womega_{11}=(\omega_{11} - \omega_{12}^2/\omega_{22})$. Then,
\begin{align*}
    f(\y\mid\lambda) & = \int f(\y\mid\bOmega)f(\bOmega\mid\tau,\,\lambda)f(\tau\mid\lambda)\,d\tau\,d\bOmega\\
     & = \cbglinv\left(\frac{1}{2\pi}\right)^{n+\frac{1}{2}}\frac{\lambda^2}{4}\bigintsss \frac{\womega_{11}^\frac{n}{2}}{\sqrt{\tau}}\exp\left(\frac{-(\lambda+s_{11})\womega_{11}}{2}\right)\exp\left(-s_{12}\omega_{12}-\frac{\omega_{12}^2}{2\tau}-\frac{(\lambda+s_{11})\omega_{12}^2}{2\omega_{22}}\right)\\
     &\quad\times\omega_{22}^\frac{n}{2}\exp\left(\frac{-(\lambda+s_{22})\omega_{22}}{2}\right)f(\tau\mid\lambda)d\womega_{11}\,d\omega_{12}\,d\omega_{22}\,d\tau\\
     & = \cbglinv\frac{\lambda^2\Gamma\left(\frac{n}{2}+1\right)}{2^{\frac{n+3}{2}}\pi^{n+\frac{1}{2}}(\lambda+s_{11})^{\frac{n}{2}+1}}\bigintsss\frac{1}{\sqrt{\tau}}\exp\left(-s_{12}\omega_{12}-\frac{\omega_{12}^2}{2\tau}-\frac{(\lambda+s_{11})\omega_{12}^2}{2\omega_{22}}\right)\\
     &\quad\times \omega_{22}^\frac{n}{2}\exp\left(\frac{-(\lambda+s_{22})\omega_{22}}{2}\right)f(\tau\mid\lambda)d\omega_{12}\,d\omega_{22}\,d\tau\\
     & = \cbglinv\frac{\lambda^2\Gamma\left(\frac{n}{2}+1\right)}{2^{\frac{n}{2}+1}\pi^{n}(\lambda+s_{11})^{\frac{n}{2}+1}}\bigintsss \omega_{22}^\frac{n}{2}\exp\left(\frac{-(\lambda+s_{22})\omega_{22}}{2}\right)f(\tau\mid\lambda)\left(\frac{1}{\tau}+\frac{\lambda+s_{11}}{\omega_{22}}\right)^{-\frac{1}{2}}\\
      &\quad\times\exp\left(\frac{1}{2} s_{12}^2\left(\frac{1}{\tau}+\frac{\lambda+s_{11}}{\omega_{22}}\right)^{-1}\right)\frac{1}{\sqrt{\tau}}d\tau\,d\omega_{22}\\
      &=  \cbglinv\frac{\lambda^4\Gamma\left(\frac{n}{2}+1\right)}{2^{\frac{n+4}{2}}\pi^{n}(\lambda+s_{11})^{\frac{n}{2}+1}}\int_{0}^\frac{\omega_{22}}{\lambda+s_{11}}z^{-\frac{3}{2}}\exp\left(\frac{\omega_{22}}{2}\left[\frac{\lambda^2+s_{12}^2}{\lambda+s_{11}}\right]\right)\exp\left(-\frac{\lambda^2}{2}\left(\frac{\omega_{22}}{\lambda+s_{11}}\right)^2\frac{1}{z} - \frac{s_{12}^2 z}{2}\right)\\
      &\quad\times\int\omega_{22}^\frac{n+3}{2}\exp\left(\frac{-(\lambda+s_{22})\omega_{22}}{2}\right)dz\,d\omega_{22}\quad (\text{setting } \frac{1}{\tau}+\frac{\lambda+s_{11}}{\omega_{22}} = \frac{1}{m},\;  \frac{\omega_{22}}{\lambda+s_{11}} -m = z)\\
      & = \cbglinv\frac{\lambda^3\Gamma\left(\frac{n}{2}+1\right)}{2^{\frac{n+3}{2}}\pi^{n-\frac{1}{2}}}\bigintsss\left(\frac{\omega_{22}}{\lambda+s_{11}}\right)^\frac{n+1}{2}\exp\left(\frac{-\Big[(\lambda+s_{11})(\lambda+s_{22}) - (\lambda - |s_{12}|)^2\Big]}{2}\left(\frac{\omega_{22}}{\lambda+s_{11}}\right)\right)\\
      &\quad\times F\left(\frac{\omega_{22}}{\lambda+s_{11}}\right)\frac{1}{\lambda+s_{11}}d\omega_{22} \quad (F(\cdot) \text{ is CDF of inverse-Gaussian},\text{ evaluated at }{\omega_{22}}/{(\lambda+s_{11})}).
      \end{align*}
 Substituting $t=\omega_{22}(\lambda+s_{11})^{-1}$ and rearranging the constants, we get the desired marginal as:
\begin{align*}
  f(\y\mid\lambda)&= C_{\mathrm{BGL}}^{-1}\frac{\lambda^3 \Gamma\left(\frac{n}{2}+1\right)\Gamma\left(\frac{n+3}{2}\right)}{\pi^{n-\frac{1}{2}} \Big[(\lambda+s_{11})(\lambda+s_{22})-(\lambda - |s_{12}|)^2\Big]^{(n+3)/2}}\mathbb{E}_t\left(F(t)\right),
 \end{align*}
 where,
 \begin{align*}
     &\quad\quad\quad t  \sim \mathrm{Gamma}\left(\mathrm{shape} = \frac{n+3}{2},\,\mathrm{rate=}\frac{(\lambda+s_{11})(\lambda+s_{22})-(\lambda - |s_{12}|)^2}{2}\right)
     \end{align*}
    and,
    \begin{align*}
    &\quad\quad\quad F(t) = \Phi\Bigg[\lambda{t^{1/2}}\left(\frac{|s_{12}|}{\lambda} - 1\right)\Bigg] + \exp\left(2\lambda|s_{12}|t\right) \Phi\Bigg[-\lambda{t^{1/2}}\left(\frac{|s_{12}|}{\lambda} + 1\right)\Bigg].
\end{align*}
The constant $C_{\mathrm{BGL}}$ is:
\begin{equation*}
    C_{\mathrm{BGL}} = \int \frac{1}{\sqrt{2\pi\tau}}\exp\left(-\frac{\omega_{12}^2}{2\tau}\right)\left(\frac{\lambda}{2}\right)^2\exp\left(-\frac{\lambda}{2}(\omega_{11}+\omega_{22})\right)f(\tau\mid\lambda)d\omega_{12}\,d\omega_{11}\,d\omega_{22}\,d\tau.
\end{equation*}
We write $\womega_{11} = \omega_{11} - \omega_{12}^2/\omega_{22}$ and integrate over  $\womega_{11}$. Similarly, multiplying and dividing with $\left({1}/{\tau} + {\lambda}/{\omega_{22}}\right)^{-\frac{1}{2}}$, we integrate over $\omega_{12}$. We are left to evaluate:
\begin{equation*}
    C_{\mathrm{BGL}} = \int \frac{1}{\sqrt{\tau}}\left(\frac{1}{\tau}+\frac{\lambda}{\omega_{22}}\right)^{-1/2}\left(\frac{\lambda}{2}\right)\exp\left(-\frac{\lambda}{2}\omega_{22}\right)\frac{\lambda^2}{2}\exp\left(-\frac{\lambda^2\tau}{2}\right)d\omega_{22}\,d\tau.
\end{equation*}
Substituting $1/\tau+\lambda/\omega_{22}= 1/m$ and $\omega_{22}/\lambda-m = z$, the above integral reduces to, 
\begin{equation*}
    C_{\mathrm{BGL}}= \frac{\lambda^{3/2}}{2^2}\int\frac{\omega_{22}^{3/2}\exp(-\lambda\omega_{22}/2)}{z^{3/2}}\exp\left(-\frac{\omega_{22}^2}{2z}\right)\exp(\lambda\omega_{22}/2)\, d\omega_{22}\,dz.
\end{equation*}
Finally substituting $z^{-1} = 2y/\omega_{22}^2$ and $\lambda\omega_{22}/2 = x$, yields, 
\begin{equation*}
    C_{\mathrm{BGL}} = \int_0^\infty \sqrt{x}\int_{x}^{\infty}y^{-1/2}\exp(-y)dydx\,\approx 0.67.
\end{equation*}
\section{Proof of Proposition~\ref{GHS_marginal_p2_lemma}}
\label{proof_GHS_marginal_p2_lemma}
As in the proof of Proposition~\ref{BGL_marginal_p2_lemma}, we first derive the marginal, $f(\y\mid\lambda)$, followed by the constant $C_{\mathrm{GHS}}$. We use the normal scale mixture representation $\omega_{ij}\mid\tau_{ij},\,\lambda\sim\mathcal{N}(0,\tau_{ij}^2/\lambda^2)$ and $\tau_{ij}\sim\mathcal{C}^+(0,1)$, for the horseshoe prior. For $p=2$, we have $\tau_{12}$ as the only latent scale parameter, and denote it as $\tau$. We use the fact that if $\tau^2 \mid a \sim \mathrm{InverseGamma} (1/2,1/a)$ and $a\sim\mathrm{InverseGamma}(1/2,1)$ then marginally $\tau\sim\mathcal{C}^+(0,1)$ \citep{makalic2016samplerHS}. The marginal is $ f(\y\mid\bOmega)=\int f(\y\mid\bOmega)f(\bOmega\mid\tau,\,\lambda)f(\tau\mid\nu,\,\lambda)f(\nu\mid\lambda)\,d\nu\,d\tau\,d\bOmega$ where, 
\begin{align*}
    f(\y\mid\bOmega) & = \frac{\left(\omega_{11}\omega_{22} - \omega_{12}^2\right)^\frac{n}{2}}{(2\pi)^n}\exp\left(-\frac{1}{2}(s_{11}\omega_{11}+2s_{12}\omega_{12} + s_{22}\omega_{22})\right),\\
    f(\bOmega\mid \tau,\,\lambda) & = \frac{\lambda C(\tau,\,\lambda)^{-1}}{\tau\sqrt{2\pi}}\exp\left(-\frac{\lambda^2\omega_{12}^2}{2\tau^2}\right)\left(\frac{\lambda}{2}\right)^2\exp\left(-\frac{\lambda(\omega_{11}+\omega_{22})}{2}\right),\\
    f(\tau\mid\nu,\,\lambda)f(\nu\mid\lambda)& =  \frac{2C(\tau,\,\lambda)\cghsinv}{\tau^2\sqrt{\pi\nu}}\exp\left(-\frac{1}{\nu\tau^2}\right)\frac{1}{\nu\sqrt{\pi\nu}}\exp\left(-\frac{1}{\nu}\right).
\end{align*}
The domain of integration such that $\bOmega\in \mathcal{M}_2^{+}$ is as mentioned in Equation~\eqref{range_omega_ijs_2by2}. Define $\womega_{11}=(\omega_{11} - \omega_{12}^2/\omega_{22})$. Then,
\begin{align*}
    f(\y\mid\lambda) & = \int f(\y\mid\bOmega)f(\bOmega\mid\tau,\,\lambda)f(\tau\mid\nu,\,\lambda)f(\nu\mid\lambda)\,d\nu\,d\tau\,d\bOmega\\
     & = \cghsinv\left(\frac{1}{2\pi}\right)^{n+\frac{1}{2}}\frac{\lambda^3}{4}\bigintsss \frac{\womega_{11}^\frac{n}{2}}{\tau}\exp\left(\frac{-(\lambda+s_{11})\womega_{11}}{2}\right)\exp\left(-s_{12}\omega_{12}-\frac{\lambda^2\omega_{12}^2}{2\tau^2}-\frac{(\lambda+s_{11})\omega_{12}^2}{2\omega_{22}}\right)\\
     &\quad \times\omega_{22}^\frac{n}{2}\exp\left(\frac{-(\lambda+s_{22})\omega_{22}}{2}\right)f(\tau\mid\nu,\,\lambda)f(\nu\mid\lambda)\,d\nu\,d\tau\,d\womega_{11}\,d\omega_{12}\,d\omega_{22}\\
     & = \cghsinv\frac{\lambda^3\Gamma\left(\frac{n}{2}+1\right)}{2^{\frac{n+3}{2}}\pi^{n+\frac{1}{2}}(\lambda+s_{11})^{\frac{n}{2}+1}}\bigintsss\frac{1}{\tau}\exp\left(-s_{12}\omega_{12}-\frac{\lambda^2\omega_{12}^2}{2\tau^2}-\frac{(\lambda+s_{11})\omega_{12}^2}{2\omega_{22}}\right)\\
     &\quad\times \omega_{22}^\frac{n}{2}\exp\left(\frac{-(\lambda+s_{22})\omega_{22}}{2}\right)f(\tau\mid\nu,\,\lambda)f(\nu\mid\lambda)\,d\nu\,d\tau\,d\omega_{12}\,d\omega_{22}\\
     & = \cghsinv\frac{\lambda^3\Gamma\left(\frac{n}{2}+1\right)}{2^{\frac{n}{2}+1}\pi^{n}(\lambda+s_{11})^{\frac{n}{2}+1}}\bigintsss \omega_{22}^\frac{n}{2}\exp\left(\frac{-(\lambda+s_{22})\omega_{22}}{2}\right)\frac{f(\tau\mid\nu,\,\lambda)}{\tau}\left(\frac{\lambda^2}{\tau^2}+\frac{\lambda+s_{11}}{\omega_{22}}\right)^{-\frac{1}{2}}\\
      &\quad\times\exp\left(\frac{1}{2} s_{12}^2\left(\frac{\lambda^2}{\tau^2}+\frac{\lambda+s_{11}}{\omega_{22}}\right)^{-1}\right)f(\nu\mid\lambda)d\tau\,d\nu\,d\omega_{22}\\
      &=  \cghsinv\frac{\lambda\Gamma\left(\frac{n}{2}+1\right)}{2^{\frac{n}{2}+1}\pi^{n+1}(\lambda+s_{11})^{\frac{n}{2}+1}}\int_0^{\frac{\omega_{22}}{\lambda+s_{11}}}\exp\left(\frac{m\,s_{12}^2}{2}\right)m^{-\frac{3}{2}}\int_{0}^{\infty}\frac{1}{\nu^2}\exp\left(-\frac{1}{\nu}\left(\frac{1}{m\lambda^2} - \frac{\lambda+s_{11}}{\lambda^2\omega_{22}}+1\right)\right)\\
      &\quad\times\int\omega_{22}^\frac{n}{2}\exp\left(\frac{-(\lambda+s_{22})\omega_{22}}{2}\right)d\nu\,dm\,d\omega_{22}\quad(\text{setting }\frac{\lambda^2}{\tau^2}+\frac{\lambda+s_{11}}{\omega_{22}} = \frac{1}{m})\\
       &=  \cghsinv\frac{\lambda\Gamma\left(\frac{n}{2}+1\right)}{2^{\frac{n}{2}+1}\pi^{n+1}(\lambda+s_{11})^{\frac{n}{2}+1}}\int_0^{\frac{\omega_{22}}{\lambda+s_{11}}}\exp\left(\frac{m\,s_{12}^2}{2}\right)m^{-\frac{1}{2}}\left(m+\frac{\omega_{22} - m(\lambda +s_{11})}{\lambda^2\omega_{22}}\right)^{-1}\\
       &\quad\times \int\omega_{22}^\frac{n}{2}\exp\left(\frac{-(\lambda+s_{22})\omega_{22}}{2}\right)dm\,d\omega_{22}.
\end{align*}
Substituting $t=\omega_{22}$ and rearranging the constants, we get the desired marginal as:
\begin{equation*}
     f(\y\mid\lambda) = C_{\mathrm{GHS}}^{-1}\frac{\lambda \Gamma\left(\frac{n}{2}+1\right)\Gamma\left(\frac{n}{2}+1\right)}{\pi^{n+\frac{1}{2}} \Big[(\lambda+s_{11})(\lambda+s_{22})\Big]^{\frac{n}{2}+1}}\mathbb{E}_t\left(F(t)\right),
\end{equation*}
where, 
\begin{equation*}
    t  \sim \mathrm{Gamma}\left(\mathrm{shape} = \frac{n}{2}+1,\,\mathrm{rate=}\frac{\lambda+s_{22}}{2}\right)
\end{equation*}
and,
\begin{equation*}
    F(t) =\int_0^{\frac{t}{\lambda+s_{11}}}\exp\left(\frac{ms_{12}^2}{2}\right)m^{-1/2}\left(m+\frac{t-m(\lambda+s_{11})}{\lambda^2 t}\right)^{-1} dm.
\end{equation*}
The constant $C_{\mathrm{GHS}}$ is:
\begin{equation*}
     C_{\mathrm{GHS}}  = \int \frac{\lambda}{\tau\sqrt{2\pi}}\exp\left(-\frac{\lambda^2\omega_{12}^2}{2\tau^2}\right)\left(\frac{\lambda}{2}\right)^2\exp\left(-\frac{\lambda}{2}(\omega_{11}+\omega_{22})\right)f(\tau\mid\lambda)d\omega_{12}\,d\omega_{11}\,d\omega_{22}\,d\tau,
\end{equation*}
where, $f(\tau\mid\lambda) = (2/\pi)(1+\tau^2)^{-1}$. Write $\womega_{11} = \omega_{11} - \omega_{12}^2/\omega_{22}$ and integrate over  $\womega_{11}$. Similarly, multiply and divide by $\left({1}/{\tau} + {\lambda}/{\omega_{22}}\right)^{-\frac{1}{2}}$, and integrate over $\omega_{12}$. We are left to evaluate:
\begin{equation*}
   C_{\mathrm{GHS}} = \frac{1}{\pi\lambda}\int\sqrt{\frac{\omega_{22}}{\omega_{22}+\lambda\tau^2}}\left(\frac{1}{1+\tau^2}\right)\exp\left(-\frac{\lambda\omega_{22}}{2}\right)d\tau\,d\omega_{22}.
\end{equation*}
Substituting $\omega_{22}/\lambda = m$, we get:
\begin{equation*}
     C_{\mathrm{GHS}} = \mathbb{E}_{(\tau,\,m)}\left(\sqrt{\frac{m}{m+\tau^2}}\right)\, \text{ where }\tau\sim\mathcal{C}^+(0,1),\, m\sim\exp(1/2)\text{ and }\tau\perp m.
\end{equation*}
Monte Carlo evaluation of the above expectation gives, $ C_{\mathrm{GHS}} \approx 0.64$.
\section{Computing  \texorpdfstring{$\mathrm{IV}_p$}{4p} for G-Wishart}\label{sec:gwishart_IV}
With change of variables,
\begin{equation*}
\bbeta_{\sbt\,p}^{nb(p)}  = \bomega_{\sbt\,p}^{nb(p)}\text{ and }\gamma_{pp}  = \omega_{pp} - \bomega_{\sbt\,p}^T\bOmega_{(p-1)\times(p-1)}^{-1}\bomega_{\sbt\,p}\,  = \omega_{pp} - {\bomega_{\sbt\,p}^{nb(p)}}^T\,\Big[\bOmega_{(p-1)\times(p-1)}^{-1}\Big]^{nb(p)}\,\bomega_{\sbt\,p}^{nb(p)}\,,
\end{equation*}
the  Jacobian of transformation $\left(\bomega_{\sbt\,p}^{nb(p)}\,,\,\omega_{pp}\right) \mapsto  \left(\bbeta_{\sbt\,p}^{nb(p)},\gamma_{pp}\right)$ equals 1. Thus, the density of the induced conditional posterior $(\bbeta_{\sbt\,p}^{nb(p)},\gamma_{pp}\mid\text{ rest})$ can be written as,
\small
\begin{align}
\label{beta_gamma_Gibbs_G_Wishart}
    f\!\left(\bbeta_{\sbt\,p}^{nb(p)}\,,\,\gamma_{pp} \mid\bOmega_{(p-1)\times(p-1)},\,\y,\,\G\right)  & = \mathcal{N}\left(\bbeta_{\sbt\,p}^{nb(p)}\,\middle\vert\, -\C\Big\{\s_{\sbt\,p}^{nb(p)}\,+\,\v_{\sbt\,p}^{nb(p)}\Big\}\,,\,\C\right)\nonumber\\
    &\times \mathrm{Gamma}\left(\gamma_{pp} \mid \alpha+n/2+1,\,(s_{pp}+\sv_{pp})/{2}\right),
\end{align}
\normalsize
where $\C= \left[(s_{pp}+\sv_{pp})\left[\bOmega_{(p-1)\times(p-1)}^{-1}\right]^{nb(p)}\right]^{-1}$. Thus, analogous to Equation~\eqref{eq:betagamma}, we have Equation~\eqref{beta_gamma_Gibbs_G_Wishart} which can be used to sample from the posterior of $(\bOmega\mid \y,\,\G)$ via a Gibbs sampler, by cycling over all $p$ columns and $\hat{f}\!\left({\left[\bomega_{\sbt\,p}^{*}\right]}^{\,nb(p)}\mid \y,\,\G\right)$ can be evaluated analogous to Equation~\eqref{first_gibbs_4_gibbs} as, 
\begin{align}
\label{eval_IV_p_part_1_G_Wishart}
    \hat{f}\!\left({\Big[\bomega_{\sbt\,p}^{*}\Big]}^{\,nb(p)}\mid \y,\,\G\right)  = M^{-1}\sum_{i=1}^{M} \mathcal{N}\left({\Big[\bomega_{\sbt\,p}^{*}\Big]}^{\,nb(p)} \,\middle\vert\,  -\C^{(i)}\Big[\s_{\sbt\,p}^{nb(p)}\,+\,\v_{\sbt\,p}^{nb(p)}\Big]\,,\,  \C^{(i)}\right),
    \end{align}
where $\C^{(i)}$ is the $i^\text{th}$ MCMC sample of $\C$ defined in Equation~\eqref{beta_gamma_Gibbs_G_Wishart} and ${\left[\bomega_{\sbt\,p}^{*}\right]}^{\,nb(p)}$ is a summary statistic (we use the sample average) based on the same MCMC runs. As in Wishart, we need a second restricted sampler which samples entries of $\bOmega$, with non-zero entries in $\bomega_{\sbt\,p}\,$ fixed at ${\left[\bomega_{\sbt\,p}^{*}\right]}^{\,nb(p)}$. This second sampler is used to evaluate $\hat{f}\!\left(\omega_{pp}^*\,\middle\vert\, {\left[\bomega_{\sbt\,p}^{*}\right]}^{\,nb(p)}\,,\,\y,\,\G\right)$.  Using Schur formula  $|\bOmega| = |\omega_{pp}||\wbOmega_{(p-1)\times(p-1)}|$ and the right hand side in Equation~\eqref{splitting_indicator_G_wishart}, we can write the induced conditional posterior,
\small
\begin{align*}
     \hspace{-0.5cm}f\!\left(\wbOmega_{(p-1)\times(p-1)}\,\middle\vert\, {\Big[\bomega_{\sbt\,p}^{*}\Big]}^{\,nb(p)}\,,\omega_{pp},\,\y,\,\G\right) &\propto |\wbOmega_{(p-1)\times(p-1)}|^{\alpha+n/2}\exp\left(-\frac{1}{2}\text{tr}\left(\Big[\S_{(p-1)\times(p-1)}+\V_{(p-1)\times(p-1)}\Big]\wbOmega_{(p-1)\times(p-1)}\right)\right)\\
    &\quad\times \Ind\Bigg[\wbOmega_{(p-1)\times(p-1)} +  {\Big[\bomega_{\sbt\,p}^{*}\Big]}^{\,nb(p)}\,\omega_{pp}^{-1}\,{{\Big[\bomega_{\sbt\,p}^{*}\Big]}^{\,nb(p)}}^T\,\in \mathcal{M}^+\!\left(\G_{(p-1)\times(p-1)}\right)\Bigg].
\end{align*}
\normalsize
Again using the Schur formula,
\begin{equation*}
|\wbOmega_{(p-1)\times(p-1)}| = |\wbOmega_{(p-2)\times(p-2)}||\womega_{(p-1)(p-1)} - \wbomega_{\sbt\,(p-1)}^T\wbOmega_{(p-2)\times(p-2)}^{-1}\wbomega_{\sbt\,(p-1)}|,
\end{equation*} 
and letting $\wbbeta_{\sbt\,(p-1)} = \wbomega_{\sbt\,(p-1)}\,,\,\wgamma_{(p-1)(p-1)} =  \womega_{(p-1)(p-1)} -  \wbomega_{\sbt\,(p-1)}^T\wbOmega_{(p-2)\times(p-2)}^{-1}\wbomega_{\sbt\,(p-1)}\,$,  the conditional posterior of $(\wbbeta_{\sbt\,(p-1)},\,\wgamma_{(p-1)(p-1)}\mid \mathrm{rest})$ can be derived analogous to Equation~\eqref{beta_gamma_Gibbs_G_Wishart} as,
\small
\begin{align}
\label{temp_tilde_beta_gamma_G_Wishart}
    f(\wbbeta_{\sbt\,(p-1)},\,\wgamma_{(p-1)(p-1)}\mid \mathrm{rest}) & \propto |\wgamma_{(p-1)(p-1)}|^{\alpha+n/2}\nonumber\\
    & \times \exp\Bigg(-\frac{1}{2}\Bigg[2\left(\s_{\sbt\,(p-1)}\,+\v_{\sbt\,(p-1)}\,\right)\wbbeta_{\sbt\,(p-1)} + (s_{(p-1)(p-1)} + \sv_{(p-1)(p-1)})\wgamma_{(p-1)(p-1)}\nonumber\\
    &\quad\quad\quad\quad +  (s_{(p-1)(p-1)} + \sv_{(p-1)(p-1)})\wbbeta_{\sbt\,(p-1)}^T\wbOmega_{(p-2)\times(p-2)}^{-1}\wbbeta_{\sbt\,(p-1)}\Bigg]\Bigg)\nonumber\\
    &\times\Ind\left(\wgamma_{(p-1)(p-1)}>0\right)\times\Ind\left(\wbbeta_{\sbt\,(p-1)}^{\,\,nb(p-1)\setminus\{p\}}\neq 0\right).
\end{align}
\normalsize
We pause to make a few important observations.
\begin{enumerate}
    \item $nb(p-1)\!\setminus\!\{p\}$ denotes the set of neighbors of the node $({p-1})$ excluding the node ${p}$, as encoded by the adjacency matrix $\G$. 
    \item It is implicit that entries in $\wbbeta_{\sbt\,(p-1)}^{\,\,nb^c(p-1)\setminus\{p\}}$ are fixed. This follows from Remark~\ref{remark:wish2}. In the context of the above density,  this yields
    \begin{equation*}
        \wbbeta_{\sbt\,(p-1)}^{\,\,nb^c(p-1)\setminus\{p\}} = -\Bigg[{\Big[\bomega_{\sbt\,p}^{*}\Big]}^{\,nb(p)}\,\omega_{pp}^{-1}\,{{\Big[\bomega_{\sbt\,p}^{*}\Big]}^{\,nb(p)}}^T\Bigg]^{nb^c(p-1)\setminus\{p\}}_{\sbt\,(p-1)}.
    \end{equation*}
    \item $\wbbeta_{\sbt\,(p-1)} = \begin{bmatrix} \wbbeta_{\sbt\,(p-1)}^{\,\,nb(p-1)\setminus\{p\}} ,\; \wbbeta_{\sbt\,(p-1)}^{\,\,nb^c(p-1)\setminus\{p\}}\end{bmatrix}^T$ and only the entries of $\wbbeta_{\sbt\,(p-1)}^{\,\,nb(p-1)\setminus\{p\}}$ are free to be sampled. 
\end{enumerate}
Armed with these observations, the density $f\left(\wbbeta_{\sbt\,(p-1)}^{\,\,nb(p-1)\setminus\{p\}},\,\wgamma_{(p-1)(p-1)}\mid \mathrm{rest}\right)$ from Equation~\eqref{temp_tilde_beta_gamma_G_Wishart} can be written as, 
\small
\begin{align}
\label{tilde_beta_gamma_G_Wishart}
    f\left(\wbbeta_{\sbt\,(p-1)}^{\,\,nb(p-1)\setminus\{p\}},\,\wgamma_{(p-1)(p-1)}\mid \mathrm{rest}\right) & =\mathcal{N}\left(\wbbeta_{\sbt\,(p-1)}^{\,\,nb(p-1)\setminus\{p\}}\,\middle\vert\,-\widetilde{\C}\,\widetilde{\bmu},\,\widetilde{\C}\right)\nonumber\\
    & \times \text{Gamma}\left(\wgamma_{(p-1)(p-1)}\,\middle\vert\, \alpha+n/2+1,\,(s_{(p-1)(p-1)} + \sv_{(p-1)(p-1)})/2\right),
\end{align}
\normalsize
where, 
\small
\begin{align*}
    \widetilde{\bmu} & = \s_{\sbt\,(p-1)}^{nb(p-1)\setminus\{p\}}\, + \v_{\sbt\,(p-1)}^{nb(p-1)\setminus\{p\}}\, + \left(s_{(p-1)(p-1)} + \sv_{(p-1)(p-1)}\right)\Bigg[\Big[\wbOmega_{(p-2)\times(p-2)}^{-1}\Big]^{nb^c(p-1)\setminus\{p\}}\times\wbbeta_{\sbt\,(p-1)}^{\,\,nb^c(p-1)\setminus\{p\}}\,\Bigg], \\
    \widetilde{\C} & = \Bigg[\left(s_{(p-1)(p-1)} + \sv_{(p-1)(p-1)}\right)\Big[\wbOmega_{(p-2)\times(p-2)}^{-1}\Big]^{nb(p-1)\setminus\{p\}}\Bigg]^{-1}.
\end{align*}
\normalsize
Equation~\eqref{tilde_beta_gamma_G_Wishart} can be used to \emph{sample} from the posterior of $\left(\wbOmega_{(p-1)\times(p-1)}\mid {\Big[\bomega_{\sbt\,p}^{*}\Big]}^{\,nb(p)}\,,\omega_{pp},\,\y,\,\G\right)$ via a block Gibbs sampler, by holding the $p$th column fixed and cycling over the remaining $(p-1)$ columns. After updating all the $(p-1)$ columns of $\wbOmega_{(p-1)\times(p-1)}$ we generate the $j$th MCMC sample from  $f\left(\bOmega_{(p-1)\times(p-1)},\omega_{pp} \mid {\Big[\bomega_{\sbt\,p}^{*}\Big]}^{\,nb(p)}\,,\,\y\right)$ as, 
\begin{align*}
\bOmega^{(j)}_{(p-1)\times(p-1)}& \leftarrow \wbOmega^{(j)}_{(p-1)\times(p-1)} +  \bomega_{\sbt\,p}^*\,\bomega_{\sbt\,p}^{*T}/\omega_{pp}^{(j-1)}, \\
    \omega^{(j)}_{pp}\mid \bomega_{\sbt\,p}^*\,,\, \bOmega^{(j)}_{(p-1)\times(p-1)},\,\by_{1:p} &\sim   \mathrm{Gamma}\left(\alpha+n/2+1,\,\frac{s_{pp}+\sv_{pp}}{2}\right) +  \bomega_{\sbt\,p}^{*T}\left(\bOmega^{(j)}_{(p-1)\times(p-1)}\right)^{-1}\bomega_{\sbt\,p}^{*}.
\end{align*}
Entries corresponding to $nb^c(p)$ in $\bomega_{\sbt\,p}^{*}$ are zero (as restricted by $\G$) and the non-zero entries in $\bomega_{\sbt\,p}^*$ are equal to ${\left[\bomega_{\sbt\,p}^{*}\right]}^{\,nb(p)}$. Thus, given $\G$; $\bomega_{\sbt\,p}^*$ and ${\left[\bomega_{\sbt\,p}^{*}\right]}^{\,nb(p)}$ can be used interchangeably. Hence approximating the value of $f\!\left(\omega_{pp}^*\mid {\left[\bomega_{\sbt\,p}^{*}\right]}^{\,nb(p)}\,,\,\y,\,\G\right)$ is straightforward using Equation~\eqref{second_gibbs_4_gibbs}. With this approximation, along with Equation~\eqref{eval_IV_p_part_1_G_Wishart}, we complete the evaluation of $\mathrm{IV}_p$.

\section{Computing  \texorpdfstring{$\mathrm{III}_{p-1}, \ldots, \mathrm{III}_1$ }{all3p} for G-Wishart}\label{sec:gwishart_III}
Before generalizing about how to evaluate the term $\mathrm{III}_{j},\,j<p$, we start with $\mathrm{III}_{p-1}$ and show that it can be evaluated as a product of a Gaussian and a generalized inverse Gaussian (GIG) densities \citep[see, e.g.][]{barn77}. This holds true for terms $\mathrm{III}_{j},\,j\leq p-1$ and the term $\mathrm{III}_{1}$ is evaluated as a gamma density. Writing the conditional prior $f(\wbOmega_{(p-1)\times(p-1)}\mid\bomega_{\sbt\,p}\,,\,\omega_{pp},\,\G)$ from Equation~\eqref{G_WIshart_prior_with_indicator_split}, we obtain:
\begin{align*}
  f(\wbOmega_{(p-1)\times(p-1)}\mid\bomega_{\sbt\,p}\,,\,\omega_{pp},\,\G) & \propto |\wbOmega_{(p-1)\times(p-1)}|^\alpha\exp\left(-\frac{1}{2}\text{tr}\Big[\V_{(p-1)\times(p-1)}\wbOmega_{(p-1)\times(p-1)}\Big]\right)\times\nonumber\\
    & \quad\Ind\Big[\wbOmega_{(p-1)\times(p-1)} + \bomega_{\sbt\,p}\,\omega_{pp}^{-1}\,\bomega_{\sbt\,p}^T\,\in \mathcal{M}^+\!\left(\G_{(p-1)\times(p-1)}\right)\Big].
\end{align*}
Recalling the definition of $\wbOmega_{(p-1)\times(p-1)}$ from Equation~\eqref{Omega_tilde_p_minus_1} and using the Schur formula,
\begin{equation*}
|\wbOmega_{(p-1)\times(p-1)}| = |\wbOmega_{(p-2)\times(p-2)} -\wbomega_{\sbt\,(p-1)}\,\,\womega_{(p-1)(p-1)}^{-1}\,\,\wbomega_{\sbt\,(p-1)}^{T}||\womega_{(p-1)(p-1)}|,
\end{equation*}
the conditional prior can be written as, 
\small
\begin{align*}
 \hspace{-0.5cm} f(\wbOmega_{(p-1\times(p-1))}\mid\bomega_{\sbt\,p}\,,\,\omega_{pp},\,\G) & \propto |\womega_{(p-1)(p-1)}|^\alpha|\wbOmega_{(p-2)\times(p-2)} -\wbomega_{\sbt\,(p-1)}\,\,\womega_{(p-1)(p-1)}^{-1}\,\,\wbomega_{\sbt\,(p-1)}^{T}|^\alpha\nonumber\\
  &\times\exp\Bigg(-\frac{1}{2}\left(\text{tr}\Big[\V_{(p-2)\times(p-2)}\wbOmega_{(p-2)\times(p-2)}\Big]+2\wbomega_{\sbt\,(p-1)}^T\v_{\sbt\,(p-1)}+\sv_{(p-1)(p-1)}\womega_{(p-1)(p-1)}\right)\Bigg)\nonumber\\
    &\times\Ind\Big[\wbOmega_{(p-1)\times(p-1)} + \bomega_{\sbt\,p}\,\omega_{pp}^{-1}\,\bomega_{\sbt\,p}^T\,\in \mathcal{M}^+\!\left(\G_{(p-1)\times(p-1)}\right)\Big].
\end{align*}
\normalsize
Following the update from Algorithm~\ref{algo_term_1_extra_details}, $\wbOmega_{(p-2)\times(p-2)}\leftarrow \wbOmega_{(p-2)\times(p-2)} -\wbomega_{\sbt\,(p-1)}\,\,\womega_{(p-1)(p-1)}^{-1}\,\,\wbomega_{\sbt\,(p-1)}^{T}\,$, the conditional prior density of $f(\wbomega_{\sbt\,(p-1)}\,,\,\womega_{(p-1)(p-1)}\mid \text{rest})$ can be obtained as, 
\small
\begin{align*}
  f(\wbomega_{\sbt\,(p-1)}\,,\,\womega_{(p-1)(p-1)}\mid \text{rest}) & \propto
  |\womega_{(p-1)(p-1)}|^\alpha|\times\exp\Bigg(-\frac{1}{2}\Bigg(2\wbomega_{\sbt\,(p-1)}^T\v_{\sbt\,(p-1)}
  +\sv_{(p-1)(p-1)}\womega_{(p-1)(p-1)}\\
 &  + \wbomega_{\sbt\,(p-1)}^{T}\,\V_{(p-2)\times(p-2)}\,\womega_{(p-1)(p-1)}^{-1}\,\,\wbomega_{\sbt\,(p-1)} \Bigg)\Bigg)\nonumber\\
  &  \times\Ind\Big[\wbomega_{\sbt\,(p-1)}^{\,\,nb(p-1)\setminus\{p\}}\neq 0\Big]\times\Ind(\womega_{(p-1)(p-1)}>0)  .
\end{align*}
\normalsize
We make a few observations analogous to those following Equation~\eqref{temp_tilde_beta_gamma_G_Wishart}.
\begin{enumerate}
    \item It is implicit that entries in $\wbomega_{\sbt\,(p-1)}^{\,\,nb^c(p-1)\setminus\{p\}}$ are fixed. In particular, 
    \begin{equation*}
        \wbomega_{\sbt\,(p-1)}^{\,\,nb^c(p-1)\setminus\{p\}} = -\Big[\bomega_{\sbt\,p}\,\omega_{pp}^{-1}\,{\bomega_{\sbt\,p}}^T\Big]^{nb^c(p-1)\setminus\{p\}}_{\sbt\,(p-1)}.
    \end{equation*}
    \item $\wbomega_{\sbt\,(p-1)} = \begin{bmatrix} \wbomega_{\sbt\,(p-1)}^{\,\,nb(p-1)\setminus\{p\}}, \;  \wbomega_{\sbt\,(p-1)}^{\,\,nb^c(p-1)\setminus\{p\}}\end{bmatrix}^T$ and to evaluate $\mathrm{III}_{p-1}$, we need the density on the entries of $\wbomega_{\sbt\,(p-1)}^{\,\,nb(p-1)\setminus\{p\}}$.
\end{enumerate}
With these, the conditional prior density of $f\left(\wbomega_{\sbt\,(p-1)}^{\,\,nb(p-1)\setminus\{p\}},\,\womega_{(p-1)(p-1)}\mid \mathrm{rest}\right)$ can be written as, 
\begin{align}
\label{G_Wishart_III_p_minus_1}
    f\left(\wbomega_{\sbt\,(p-1)}^{\,\,nb(p-1)\setminus\{p\}},\,\womega_{(p-1)(p-1)}\mid \mathrm{rest}\right) & = \mathcal{N}\left(\wbomega_{\sbt\,(p-1)}^{\,\,nb(p-1)\setminus\{p\}}\,\middle\vert\, -\widetilde{\U}\widetilde{\bzeta},\,\widetilde{\U}\right)\times\text{GIG}\left(\womega_{(p-1)(p-1)}\,\mid\, a,\,b,\,q\right). 
\end{align}
In Equation~\eqref{G_Wishart_III_p_minus_1}, $\text{GIG}(x\mid a,\,b,\,q)$ denotes a generalized inverse Gaussian, with density $f(x) = \{{(a/b)^{q/2}}/({2K_{q}(\sqrt{ab})})\}x^{q-1}\exp\left\{-({1}/{2})\left(ax+{b}/{x}\right)\right\}$, where $K_q(\cdot)$ denotes modified Bessel function of the second kind and, 
\begin{align*}
    \widetilde{\U} =\Bigg[\womega_{(p-1)(p-1)}^{-1}\V_{(p-2)\times(p-2)}^{nb(p-1)\setminus\{p\}}\Bigg]^{-1},\,\widetilde{\bzeta} =  \v_{\sbt\,(p-2)}^{nb(p-1)\setminus\{p\}} + \womega_{(p-1)(p-1)}^{-1}\,\V_{(p-2)\times(p-2)}^{nb^c(p-1)\setminus\{p\}}\,\,\wbomega_{\sbt\,(p-1)}^{\,\,nb^c(p-1)\setminus\{p\}},\\
    a = \sv_{(p-1)(p-1)},\,b = \Bigg[\wbomega_{\sbt\,(p-1)}^{\,\,nb^c(p-1)\setminus\{p\}}\Bigg]^T\, \wbomega_{\sbt\,(p-1)}^{\,\,nb^c(p-1)\setminus\{p\}},\,q = \alpha + \frac{|nb(p-1)\!\setminus\!\{p\}|}{2}+1.
\end{align*}
Similarly, terms $\mathrm{III}_{p-2},\ldots,\mathrm{III}_2$ can be evaluated as products of Gaussian and generalized inverse Gaussian densities. Finally, the conditional prior density of $\womega_{11}$ is, $f(\womega_{11}\mid\text{rest})\propto \womega_{11}^\alpha\exp(-\sv_{11}\womega_{11}/2)$, with the definition of $\womega_{11}$ available from Algorithm~\ref{algo_term_1_extra_details}. Hence, the conditional prior density on $\womega_{11}$ is $\text{Gamma}(\alpha+1,\; \sv_{11}/2)$.

\section{Computing  \texorpdfstring{$\mathrm{IV}_{p-1}, \ldots, \mathrm{IV}_{1}$}{all4p} for G-Wishart}\label{sec:gwishart_IVp}
For $\mathrm{IV}_{p-1}$, we need to evaluate,
\begin{equation*}
f\left(\left[\wbomega_{\sbt\,(p-1)}^{*}\right]^{\,\,nb(p-1)\setminus\{p\}},\,\womega_{(p-1)(p-1)}^{*}\,\middle\vert\,\Big[\bomega_{\sbt\,p}^{*}\Big]^{nb(p)},\,\omega_{pp}^{*}\,,\,\y_{1:(p-1)}\right).
\end{equation*}
Here $f\left(\left[\wbomega_{\sbt\,(p-1)}^{*}\right]^{\,\,nb(p-1)\setminus\{p\}}\,\middle\vert\,\left[\bomega_{\sbt\,p}^{*}\right]^{nb(p)},\,\omega_{pp}^{*}\,,\,\y_{1:(p-1)}\right)$ can be approximated using the normal density in Equation~\eqref{tilde_beta_gamma_G_Wishart} with two caveats: (a) $\omega_{pp}$ is fixed at $\omega_{pp}^*$ and is not updated while sampling $\wbOmega_{(p-1)\times(p-1)}$, as in the updates following Equation~\eqref{tilde_beta_gamma_G_Wishart} and (b) the sample covariance matrix corresponds to that of $\y_{1:(p-1)}$. Next, we need a restricted second sampler which updates $\wbOmega_{(p-2)\times(p-2)}$ (Algorithm~\ref{algo_term_1_extra_details}) with $\Big[\wbomega_{\sbt\,(p-1)}\Big]^{nb(p-1)\setminus\{p\}}$ fixed. This restricted sampler is used to approximate $f\left(\womega_{(p-1)(p-1)}^{*}\,\middle\vert\,\Big[\wbomega_{\sbt\,(p-1)}^{*}\Big]^{\,\,nb(p-1)\setminus\{p\}}\,,\,\Big[\bomega_{\sbt\,p}^{*}\Big]^{nb(p)},\,\omega_{pp}^{*}\,,\,\y_{1:(p-1)}\right)$. The details for this, and the calculations for $\mathrm{IV}_{p-2}, \ldots, \mathrm{IV}_1$  are similar to the calculations for $\mathrm{IV}_p$ with appropriate adjustments to the Schur complement, and are omitted.

\section{MCMC Diagnostics for Chib in BGL and GHS}\label{sec:diag}
\begin{figure}[!htb]
    \centering
    \includegraphics[width=\textwidth]{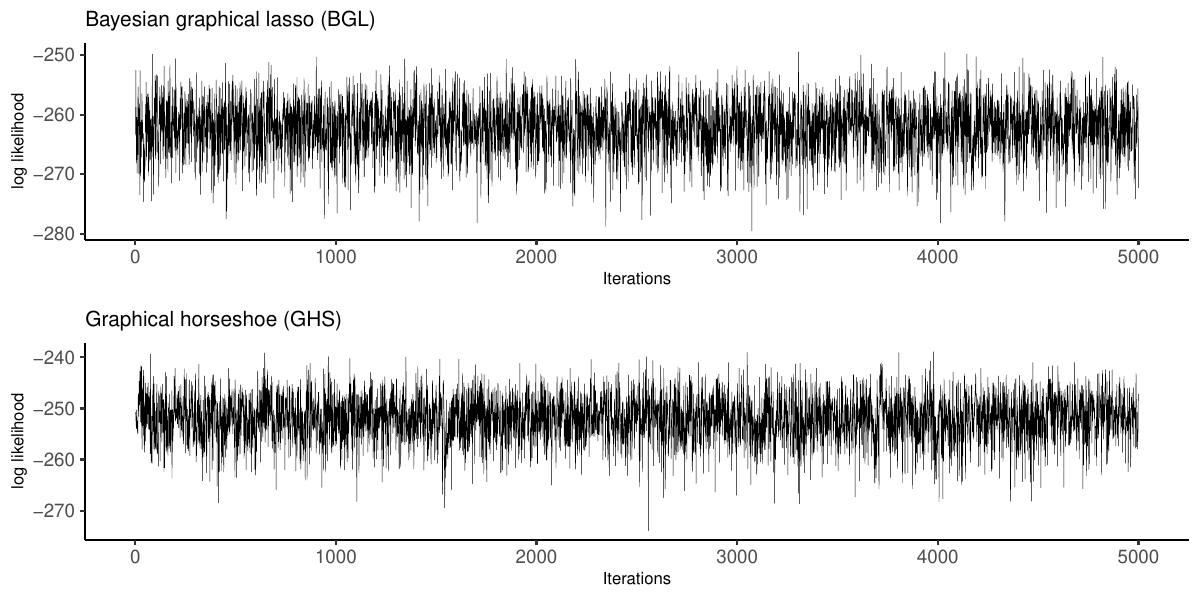}
    \caption{Trace plots of log likelihood vs. index of saved posterior samples for Bayesian graphical lasso (BGL) -- top panel and graphical horseshoe (GHS) -- bottom panel, when $p=10,\,\lambda=2\text{ and }n=20$. The log likelihood is considered at the first row of the telescoping sum, when elements in all rows (columns) of $\bOmega$ are sampled.}
    \label{trace_plots_BGL_GHS_sample}
\end{figure}

Since success of Chib's method depends on an underlying valid Gibbs sampler, we provide representative diagnostic plots for BGL and GHS in Figure~\ref{trace_plots_BGL_GHS_sample}, indicating good mixing. The plots for other dimensions and settings are similar. 
\section{Computation Times for the Competing Procedures}
We list the computational times for the competing procedures for the simulation settings described in Sections~\ref{sec:wishart},~\ref{sec:element} and~\ref{sec:gwishart} in Supplementary Tables \ref{Wishart_times_supp}$-$\ref{G_Wishart_times_banded_supp}.

\begin{table}[!tbh]
\centering
\footnotesize
\begin{tabular}{c>{\color{black}}cccc}
\hline
Dimension and Parameters & Proposed & AIS  & Nested & HM\\ \hline
$(p =5,\,n=10,\,\alpha=7)$ & 0.18        &  21.15  &  0.62    &  2.06 \\ 
$(p =10,\,n=20,\,\alpha=13)$ & 0.46    &  33.19   &   0.81   &  3.33 \\ 
$(p =15,\,n=30,\,\alpha=20)$ & 1.02   &  47.87   &   0.89   &  4.73 \\ 
$(p =25,\,n=50,\,\alpha=33)$ & 3.04   & $-$   &  $-$  & 9.87 \\
$(p =30,\,n=60,\,\alpha=45)$ & 4.81   & $-$ &   $-$   & 13.28  \\
$(p =40,\,n=80,\,\alpha=70)$ & 11.34   & $-$ &   $-$   & 26.27  \\
$(p =50,\,n=75,\,\alpha=100)$ & 21.44    & $-$    &  $-$   &  42.92 \\
$(p =100,\,n=150,\,\alpha=200)$ & 211.7    & $-$    &  $-$   &  247.91 \\
$(p =125,\,n=175,\,\alpha=250)$ & 474.65    & $-$    &  $-$   &  485.16 \\
\hline
\end{tabular}
\normalsize
\caption{Average computational time of competing approaches, in seconds, for estimation of the marginal likelihood in Wishart (see Table~\ref{wishart_result_table}). Computational times for AIS~\citep{neal2001annealed} and Nested~\citep{skilling2006nested} are not presented in the cases when the estimate of the marginal likelihood is $-\infty$.}
\label{Wishart_times_supp}
\end{table}

\begin{table}[!tbh]
\centering
\footnotesize
\begin{tabular}{c>{\color{black}}cccc}
\hline
Dimension and Parameters & Proposed  & AIS  & Nested & HM \\ \hline
 $(p =2,\, \lambda = 0.4,\, n=4)$ & 0.02 &  51.02   & 1.34  & 0.43  \\
 $(p =2,\, \lambda = 1,\, n=5)$   &  0.02 & 51.02   & 1.34  & 0.43   \\
 $(p =2,\, \lambda = 2,\, n=10)$  &  0.02 & 51.02   & 1.34  & 0.43   \\
 $(p =5,\, \lambda = 1,\, n=10)$  &  0.34 & 204.55   & 5.44  & 2.29      \\
 $(p =10,\, \lambda = 2,\, n=20)$ &  2.57 & 321.12   & 8.76  & 4.45    \\
 $(p =15,\, \lambda = 3,\, n=30)$  & 8.53 & $-$   & $-$  & 7.81    \\
 $(p =25,\, \lambda = 5,\, n=50)$  & 40.72 & $-$   & $-$   &  17.4   \\
 $(p =30,\, \lambda = 6,\, n=60)$  & 71.57 &  $-$    & $-$ &   24.1 \\
 $(p =40,\, \lambda = 175,\, n=90)$ & 182.51 & $-$    &  $-$ & 41.00   \\
  $(p =50,\, \lambda = 140,\, n=130)$ & 374.18 & $-$    &  $-$ & 65.96    \\
\hline
\end{tabular}
\normalsize
\caption{Average computational time of competing approaches, in seconds, for estimation of the marginal likelihood  in Bayesian graphical lasso (BGL, see Table~\ref{BGL_results_compact}). Computational times for AIS~\citep{neal2001annealed} and Nested~\citep{skilling2006nested} are not presented in the cases when the estimate of the marginal likelihood is $-\infty$.}
\label{BGL_times_supp}
\end{table}

\begin{table}[!tbh]
\centering
\footnotesize
\begin{tabular}{c>{\color{black}}cccc}
\hline
Dimension and Parameters & Proposed & AIS  & Nested & HM \\ \hline
 $(p =2,\, \lambda = 0.4,\, n=4)$ & 0.02 & 70.09    &  2.04 &   0.61 \\
 $(p =2,\, \lambda = 1,\, n=5)$   & 0.02 & 70.09     & 2.04  &   0.61  \\
 $(p =2,\, \lambda = 2,\, n=10)$  & 0.02 & 70.09     & 2.04  &   0.61 \\
 $(p =5,\, \lambda = 1,\, n=10)$  &  0.31 & 264.08   & 8.15  &   3.16  \\
 $(p =10,\, \lambda = 2,\, n=20)$ &  2.14 & 474.41   & 16.08  &   5.22  \\
 $(p =15,\, \lambda = 3,\, n=30)$  & 6.94 & $-$    & 51.13  &  8.58   \\
 $(p =25,\, \lambda = 5,\, n=50)$  & 32.64 & $-$      & $-$   & 16.95    \\
 $(p =30,\, \lambda = 6,\, n=60)$  & 57.33 & $-$     &  $-$  &  22.02   \\
  $(p =40,\, \lambda = 140,\, n=90)$ & 147.56  & $-$    &  $-$ & 35.38 \\
  $(p =50,\, \lambda = 190,\, n=120)$ &  305.51  &$-$    & $-$  & 55.92   \\
\hline
\end{tabular}
\normalsize
\caption{Average computational time of competing approaches, in seconds, for estimation of the marginal likelihood in graphical horseshoe (GHS, see Table~\ref{GHS_results_compact}). Computational times for AIS~\citep{neal2001annealed} and Nested~\citep{skilling2006nested} are not presented in the cases when the estimate of the marginal likelihood is $-\infty$. }
\label{GHS_times_supp}
\end{table}

\begin{table}[!h]
\centering
\footnotesize
\begin{tabular}{c>{\color{black}}ccccc}
\hline
Dimension and Parameters & Proposed   & AKM & AIS  & Nested & HM \\ \hline
 $(p =5,\, \alpha = 2,\, n=10)$  & 0.15 & 0.031 & 222.63 & 8.57 & 3.71 \\
 $(p =10,\, \alpha = 3,\, n=20)$ & 1.26 &  0.094 & 544.07 & 8.29  & 16.86   \\
 $(p =15,\, \alpha = 5,\, n=30)$ & 4.37 & 0.33 & 865.67 & 24.16 & 15.12   \\
 $(p =25,\, \alpha = 10,\, n=50)$ & 21.04 & 1.83 & $-$  & $-$   & 26.57    \\
 $(p =30,\, \alpha = 20,\, n=60)$ & 39.66 & 3.66 & $-$  & $-$  & 34.39   \\
 $(p =40,\, \alpha = 25,\, n=80)$ & 96.17 & 9.86 & $-$  & $-$  & 52.64   \\
 $(p =50,\, \alpha = 15,\, n=100)$ & 196.48 & 25.26 & $-$  & $-$  & 80.72   \\
 $(p =100,\, \alpha = 50,\, n=200)$ & 2464.91 & $-$ & $-$  & $-$  &  341.23  \\
 $(p =125,\, \alpha = 50,\, n=250)$ & 5572.73 & $-$ & $-$  & $-$  &  833.19  \\
\hline
\end{tabular}
\normalsize
\caption{Average computational time of competing approaches, in seconds, for estimation of the marginal likelihood  in G-Wishart (see Table~\ref{G_Wishart_results_compact}). Computational times for AKM~\citep{Atay05}, AIS~\citep{neal2001annealed} and Nested~\citep{skilling2006nested} are not presented in the cases when the estimate of the marginal likelihood is $-\infty$.}
\label{G_Wishart_times_supp}
\end{table}

\begin{table}[!h]
\color{black}
\centering
\footnotesize
\begin{tabular}{cccccc}
\hline
Dimension and Parameters & Proposed   & AKM & AIS  & Nested & HM \\ \hline
 $(p =5,\, \alpha = 2,\, n=10)$  & 0.21 & 0.02 & 102.18 &  3.84& 2.94 \\
 $(p =10,\, \alpha = 3,\, n=20)$ & 0.94 &  0.09 & 202.69 & 6.19& 5.86   \\
 $(p =15,\, \alpha = 5,\, n=30)$ & 2.58 & 0.24 &  $-$ & 9.03 & 8.13  \\
 $(p =25,\, \alpha = 10,\, n=50)$ & 9.93 & 1.38 & $-$  & $-$   & 15.56    \\
 $(p =30,\, \alpha = 20,\, n=60)$ & 16.8 & 2.83 & $-$  & $-$  & 20.84  \\
 $(p =40,\, \alpha = 25,\, n=80)$ & 41.16 & 8.96 & $-$  & $-$  & 30.11   \\
 $(p =50,\, \alpha = 15,\, n=100)$ & 85.67 & 21.87 & $-$  & $-$  & 46.05   \\
 $(p =100,\, \alpha = 50,\, n=200)$ & 1063.76 & $-$ & $-$  & $-$  &  241.96  \\
 $(p =125,\, \alpha = 50,\, n=250)$ & 2455.3 & $-$ & $-$  & $-$  &  668.67  \\
\hline
\end{tabular}
\normalsize
\captionsetup{labelfont={color=black}, textfont={color=black}}
\caption{Average computational time of competing approaches, in seconds, for estimation of the marginal likelihood  in G-Wishart, banded tri-diagonal $\G$ (see Table~\ref{G_Wishart_results_compact_banded}). Computational times for AKM~\citep{Atay05}, AIS~\citep{neal2001annealed} and Nested~\citep{skilling2006nested} are not presented in the cases when the estimate of the marginal likelihood is $-\infty$.}
\label{G_Wishart_times_banded_supp}
\end{table}
\end{document}